\documentclass[prb,reprint,twocolumn,amsmath,amssymb,floatfix]{revtex4-2}
\usepackage{graphicx,comment,color}

\usepackage{subcaption}
\usepackage{physics}
\usepackage{amsmath}
\usepackage{eufrak}
\usepackage{natbib}

\begin{document}

\title{Counter-diabatic driving in the classical $\beta$-Fermi-Pasta-Ulam-Tsingou chain}
\author{Nik O. Gjonbalaj, David K. Campbell, Anatoli Polkovnikov }
\affiliation{Department of Physics, Boston University, Boston, Massachusetts 02215, USA}

\date{\today}

\begin{abstract}
	Shortcuts to adiabaticity (STAs) have been used to make rapid changes to a system while eliminating or minimizing excitations in the system’s state. In quantum systems, these shortcuts allow us to minimize inefficiencies and heating in experiments and quantum computing protocols, but the theory of STAs can also be generalized to classical systems. We focus on one such STA, approximate counter-diabatic (ACD) driving, and numerically compare its performance in two classical systems: a quartic anharmonic oscillator and the $\beta$ Fermi-Pasta-Ulam-Tsingou lattice. In particular, we modify an existing variational technique to optimize the approximate driving and then develop classical figures of merit to quantify the performance of the driving. We find that relatively simple forms for the ACD driving can dramatically suppress excitations regardless of system size. ACD driving in classical nonlinear oscillators could have many applications, from minimizing heating in bosonic gases to finding optimal local dressing protocols in interacting field theories.
\end{abstract}

\maketitle

\section{INTRODUCTION}

Losses and noise represent a ubiquitous challenge in the control and use of any physical system. Whether we are trying to prepare a high-fidelity quantum state for some calculation, maximizing the efficiency of some thermodynamic process, or tuning some parameter of a sensitive Hamiltonian, transitions away from the desired outcome are inevitable. Adiabatic evolution represents the most direct method for combating these inefficiencies, as asymptotically slow evolution removes all diabatic excitations. However, during the resulting long time scales, the system suffers from environmental noise and decoherence, making true adiabatic transport not only costly but impossible to implement. This tension has led to the development of many ``shortcuts to adiabaticity'' (STAs) \cite{STAs_Chen,Gu_ry_Odelin_2019,STAs_Campo1,del_Campo_2019} that achieve exactly or nearly the same effect as adiabatic transport in a fraction of the time, thereby limiting the effect of outside forces and increasing the speed of the process. These methods have shown great promise theoretically and experimentally and apply to nearly any system where high-fidelity control is desired.

One such STA that has found application in many different systems is counter-diabatic (CD) or transitionless driving \cite{demirplak_2003,demirplak_2005,Berry_2009}. This method adds to the bare Hamiltonian a driving counterterm constructed out of the ``adiabatic gauge potential'' (AGP). In quantum geometry, this object reveals how eigenstates are connected to each other as the control parameter in the Hamiltonian defining the adiabatic transformation is tuned. This counterterm kills off all diabatic transitions exactly for any tuning speed. In simple systems, CD driving offers a powerful method for exactly transporting eigenstates as the Hamiltonian is altered, but in many-body systems this approach breaks down, and the AGP generically does not exist as a local continuous operator. Recently, however, Ref.~\cite{Sels_2017} found that in many cases an approximate local variational AGP exists, which allows one to significantly suppress dissipative effects even in chaotic systems. This method has been extensively applied to quantum spin systems with many promising results \cite{Kolodrubetz_2017,Hartmann_2020,Hartmann_2020_2,Prielinger_2021,Claeys_2019,kumar2021counterdiabatic}. Moreover, Ref.~\cite{Claeys_2019} showed that this ACD protocol can be realized using Floquet engineering with only the Hamiltonian and the control field.

Although ACD driving can easily be generalized to classical spins, the method outlined in the original paper \cite{Sels_2017} does not apply to systems with unbounded local Hilbert spaces. As such, the original procedure is limited in scope to spin chains and fermionic systems, leaving out coupled oscillators and collections of bosons. In this work, we fill in the gaps and generalize ACD driving to systems with unbounded Hilbert spaces of any dimension. In particular, we focus on coupled nonlinear oscillators, where the classical phase space and local quantum Hilbert space are unbounded. Our formalism and the developed protocols apply equally to quantum and classical systems. However, in order to verify the performance of the ACD protocol, we have to rely on numerical simulations, which are currently out of the reach for quantum systems with many degrees of freedom. In small systems we are able to show that quantum and classical ACD protocols perform very similarly. We begin by working with a single anharmonic oscillator to explain the procedure before moving to a more complicated system: the $\beta$ Fermi-Pasta-Ulam-Tsingou (FPUT) lattice. This nonlinear chain not only provides a simple many-body Hamiltonian to study
but also has many deep connections to statistical mechanics and thermalization. Originally, Fermi, Pasta, Ulam, and Tsingou investigated this system numerically with the expectation that the nonlinearity would cause the chain to thermalize quickly and reach equipartition \cite{fput_original}. However, when they initialized the system with all the energy in the first normal mode, they famously discovered that almost all the energy returned to this initial mode in what are now called FPUT ``recurrences.'' In contradiction with the expectation of equipartition, these recurrences continue as the simulation progresses, keeping the system in a ``metastable state'' before finally thermalizing after very long time scales \cite{metastable,Bambusi_2008,Danieli_2017}. Recently, it was discovered that recurrences themselves oscillate in a periodic fashion as well, a feature dubbed ``higher-order recurrences" ~\cite{Pace_2019,Pace_Reiss_2019}. 


This intimate connection between the FPUT system, thermalization, and chaotic dynamics also motivates an application of ACD driving. Indeed, it was recently shown that the AGP can act as an extremely sensitive probe of chaos in certain quantum systems \cite{Pandey_2020}. Developing approximate AGPs for the $\beta$ FPUT lattice in future work could therefore reveal why the system remains quasi-periodic for certain initial conditions. In addition, FPUT-like recurrences come up in many theoretical \cite{NLSE1,NLSE2,hsueh2018thermalization,Balasubramanian2014HolographicTS} and experimental \cite{BEC_FPUT,Feedback_ring_FPUT} models. ACD driving can be used to efficiently prepare such systems in different nearly stationary initial conditions, where these recurrences are expected to be strongly suppressed, to study the effects of long-time thermalization.

The paper is organized as follows. We first introduce the concept of ACD driving in Section \ref{sec:acd} and discuss its subtleties before explaining our modifications to the variational method. We then apply the procedure to a simple anharmonic oscillator in Section \ref{sec:nlo} as an example implementation. In Section \ref{sec:bfput}, we apply the same machinery to the $\beta$-FPUT lattice and show how our results survive in the thermodynamic limit. Finally, we discuss instabilities that arise when one considers long-wavelength initial states like those in the original FPUT problem. We argue that these instabilities arise because the corresponding initial distribution is far from any statistical equilibrium and is very fragile against adding nonlinearity to the system. Nevertheless, even in this situation we provide a method for combating this instability by adding fluctuations to the initial state and variationally optimizing the AGP over a broader range of phase space than the actual initial distribution. We show that in this way we can stabilize the variational method, albeit at the expense of reducing its accuracy somewhat.


\section{APPROXIMATE COUNTER-DIABATIC DRIVING}

\label{sec:acd}

The concept of CD driving is most easily explained in a quantum sense, so we will begin there and then move to the classical regime. Consider a Hamiltonian $H_0(\beta(t))$ dependent on some parameter $\beta(t)$ that varies in time, and let the system start in some stationary state $\ket{\psi_i}$ at time $t=0$. We now want to ramp $\beta$ from some initial value $\beta(t=0) = \beta_i$ to some final value $\beta(t=\tau) = \beta_f$ over some ``turn-on time'' $\tau$. Assuming there are no degeneracies on the way, if we take an infinite time to ramp up the perturbation ($\tau \rightarrow \infty$), the system will undergo adiabatic evolution, finishing in a new stationary state $\ket{\psi_{\mathrm{target}}}$ of the final Hamiltonian $H_0(\beta(\tau))$. However, for any finite value of $\tau$, the system will inevitably experience diabatic transitions away from this state. The origin of these transitions is easy to understand by going to the moving frame defined by the time-dependent instantaneous basis of $H_0(\beta)$. Here, the effective moving Hamiltonian picks up an extra term familiar e.g. from doing a Galilean transformation or going to a rotating frame:
\begin{align}
	H_0^{\mathrm{eff}} = {H}_0 - \dot{\beta} {\mathcal{A}}_{\beta} ,
	\label{moving frame H}
\end{align}
where ${\mathcal{A}}_{\beta}$ is the adiabatic gauge potential (AGP), the generator of perturbations in $\beta$. Because the Hamiltonian $H_0$ is diagonal in its own instantaneous basis, all transitions from our initial stationary state occur due to the presence of the second term. Exact CD driving provides a method for exactly reaching our target state $\ket{\psi_{\mathrm{target}}}$ by evolving under a CD Hamiltonian:
\begin{align}
	H_{\mathrm{CD}}(t) = H_0(\beta(t)) + \dot{\beta}(t) \mathcal{A}_{\beta}(\beta(t)) .
\end{align}
In the moving frame, we can see that the two terms with the AGP will cancel, making the effective quantum Hamiltonian fully diagonal and eliminating transitions. It is straightforward to show that, as the generator conjugate to changes in $\beta$, the AGP satisfies \cite{Kolodrubetz_2017}
\begin{align}
	\bra{m}\mathcal{A}_{ \beta}\ket{n} = i \hbar \frac{\bra{m}\partial_{ \beta} H_0\ket{n}}{E_{n}-E_{m}} ,
	\label{pert form of agp}
\end{align}
or, equivalently,
\begin{align}
	\label{agp op eqn}
	\left[H_0, i \hbar \partial_{ \beta} H_0-\left[\mathcal{A}_{ \beta}, H_0\right]\right]=0 .
\end{align}


In simple systems, we can solve Eq.~ \eqref{agp op eqn} to find the exact AGP and implement transitionless driving. However, one can show that in general, the AGP blows up exponentially with the system size in chaotic systems ~\cite{Pandey_2020}. More specifically, quantum chaotic systems suffer from small denominators in Eq.~ \eqref{pert form of agp} as a result of the eigenstate thermalization hypothesis (ETH)~\cite{Kolodrubetz_2017}. Intuitively, this blow-up occurs because we are trying to follow eigenstates that, according to the ETH, are essentially random vectors and therefore require fine-tuning. Instead of exactly canceling off transitions for this exponentially difficult cost, we want to suppress transitions as much as possible, i.e. maximize the fidelity of our evolved state with the target state. 



Reference \cite{Sels_2017} outlines a method for finding an approximate form of the AGP in what is known as approximate CD (ACD) driving. First, we choose some ansatz for the form of the AGP, ideally one that doesn't suffer from the non-local terms typically present in chaotic systems ~\cite{Kolodrubetz_2017,Pandey_2020}. Once we have our ansatz, we can compute an action that will optimize the free parameters of the AGP ansatz as functions of $\beta$. Consider the object
\begin{align}
	G_{\beta} \left(\mathcal{A}_{\beta}^*\right) \equiv \partial_{\beta} H_0 + \frac{i}{\hbar} \left[ \mathcal{A}_{\beta}^*, H_0 \right] ,
	\label{G def}
\end{align}
where $\mathcal{A}_{\beta}^*$ is our ansatz. It can be shown that Eq.~\eqref{agp op eqn} follows from minimizing the following action $\mathcal{S}$ with respect to the all possible operators $\mathcal A_\beta^\ast$ \cite{Kolodrubetz_2017}:
\begin{align}
	\mathcal{S}(\mathcal{A}_\beta^*) = \left\langle G^2_{\beta} \left(\mathcal{A}_{\beta}^*\right) \right\rangle - \left\langle G_{\beta} \left(\mathcal{A}_{\beta}^*\right) \right\rangle^2 ,
	\label{action def}
\end{align}
where the brackets denote an average with respect to some equilibrium density matrix $\rho$ commuting with the Hamiltonian $H_0$. If the variational ansatz is complete, i.e. it spans the operator basis, then this minimization leads to the exact AGP. If the variational manifold is restricted to, for example, a class of local operators, the minimization leads to the best variational AGP within this class. One can use flexibility in choosing the equilibrium density matrix $\rho$ to target specific states of interest. For example, $\rho=|\psi_0\rangle\langle \psi_0|$ optimizes the AGP with respect to the ground state, which could be beneficial for quantum annealing protocols. A different choice of $\rho=I={1\over D}\sum_n |n\rangle \langle n|$, where $D$ is the Hilbert space size, optimizes the AGP with respect to all states. Other choices of $\rho$ can include e.g. finite temperature Gibbs ensembles or microcanonical ensembles targeting states in a specific energy range. Once we fix the form of $\rho$, we find our optimal approximate AGP by extremizing the action in the usual sense:
\begin{align}
	\frac{\delta \mathcal{S}(\mathcal{A}_{\beta}^*) }{\delta \mathcal{A}_{\beta}^* } = 0 .
\end{align}
If our operator $\mathcal A_\beta^\ast$ is expanded in a set of some fixed basis operators $X_j$: $\mathcal A_\beta^\ast=\sum_j \gamma_j X_j$, then variational optimization with respect to $\mathcal A_\beta^\ast$ is equivalent to minimization with respect to the set of parameters $\gamma_j$. Note that by construction the action is quadratic in $\gamma_j$ so this minimization is simply a minimization of a quadratic form. After finding the variational AGP we can then construct the approximate CD Hamiltonian
\begin{align}
	H_{\mathrm{ACD}}(t) = H_0 + \dot{\beta} \mathcal{A}_{\beta}^*,
	\label{cd ham}
\end{align}
which reduces transitions compared to the bare protocol.

\subsection*{From quantum spins to classical oscillators}

Previous work on ACD driving has focused on the infinite-temperature norm, where the average in Eq.~\eqref{action def} is taken with respect to the identity matrix and focused mostly on quantum spin systems~\cite{Sels_2017,Claeys_2019,Kolodrubetz_2017, Hartmann_2019, Passarelli_2020}. Such an infinite temperature optimization is ill-defined in systems like the FPUT model, where the local Hilbert space is unbounded and the infinite temperature state corresponds to infinite energy density. To proceed, we therefore have to choose a different $\rho$. As we are dealing with classical systems, let us also briefly discuss how the variational approach extends to them from the quantum regime.

In classical chaotic systems, exact CD driving is also impossible due to formal divergence of the AGP~\cite{Jarzynski_CM_chaos}. As such, we will turn to ACD driving for our classical systems. To make this shift, most of the standard relations between quantum and classical are involved: operators become functions and commutators become Poisson brackets, so now Eqn. \ref{G def} becomes
\begin{align}
	G_{\beta} \left(\mathcal{A}_{\beta}^*\right) \equiv \partial_{\beta} H_0 - \left\{ \mathcal{A}_{\beta}^*, H_0 \right\} .
\end{align}
The classical analog of the equilibrium density matrix $\rho$ is a stationary probability distribution $P(\vec q,\vec p)$, where $\vec q, \;\vec p$ are canonically conjugate phase space variables. The approximate AGP can again be found from the minimization of the action~\eqref{action def}, where the average is now taken with respect to the probability distribution $P_\beta$, i.e. 
\begin{equation}
\langle G_{\beta} \rangle= \int P(\vec q,\vec p,\beta)\, G_\beta (\vec q,\vec p)\; D\vec q D\vec p,
\end{equation}
and similarly for $\langle G^2_{\beta} \rangle$, where $D\vec q D\vec p$ denotes a differential volume of phase space. There is the same flexibility in choosing the stationary probability distribution $P(\vec q,\vec p)$ in classical systems as in choosing a stationary density matrix $\rho$ in quantum systems. In particular, a single quantum eigenstate averaging in a chaotic system corresponds to averaging over a microcanonical ensemble: $P_\beta(\vec q,\vec p)\propto \delta (E-H(\vec q,\vec p,\beta))$; similarly a finite temperature Gibbs density matrix corresponds to a finite temperature classical Gibbs probability distribution $P_\beta (\vec q,\vec p)\propto \exp[- H(\vec q,\vec p,\beta)/(k_b T)]$. 

It is intuitively clear that the optimal choice of the probability distribution $P(\vec q,\vec p)$ should be closest to the one adiabatically connected to the initial state. That is, if we initialize the system in some microcanonical ensemble, then $P_\beta$ should remain a microcanonical distribution with respect to the instantaneous Hamiltonian $H_(\vec q,\vec p,\beta)$ at the energy corresponding to constant initial entropy. The easiest way to generate such a distribution, which we use in this work, is by slowly evolving the initial distribution in time to some coupling $\beta$ between initial and final values. We then sample from this distribution to minimize the action~\eqref{action def} and find the approximate AGP. This procedure does not require finding the adiabatically connected distribution $P_\beta$ and has another clear advantage in that it applies both to integrable and nonintegrable systems, as it does not require the system to thermalize. In principle, one can repeat this procedure iteratively: evolve the initial state with the ACD protocol to find a better adiabatically connected state and reminimize the action to find a better AGP. We find that this reoptimization is not necessary and even the first iteration gives excellent results.

\section{SINGLE ANHARMONIC OSCILLATOR}
\label{sec:nlo}

To illustrate the approach, we first start with a single anharmonic oscillator with Hamiltonian given by
\begin{align}
	H_0 = \frac{p^2}{2} + \frac{x^2}{2} + \beta \frac{x^4}{4} ,
\end{align}
where units have been chosen to set the mass and frequency to 1. We initialize our system with a value of $\beta(t=0) = 0$ in a microcanonical distribution of energy $E_0 = 1$. We then ramp up to $\beta(t=\tau) = 1$ using the protocol
\begin{align}
	\beta(t) = \sin^2 \left( \frac{\pi}{2} \sin^2 \left( \frac{\pi t}{2 \tau} \right) \right), \quad t \in [0,\tau] ,
	\label{beta protocol}
\end{align}
from Ref. \cite{Sels_2017}. This protocol is smooth at the boundaries, suppressing spurious non-adiabatic transitions related to turning on and turning off the drive, which appear if $\dot \beta$ jumps at the protocol boundaries. Under perfect adiabatic evolution ($\tau \rightarrow \infty$), the system would end in a microcanonical distribution of the new Hamiltonian. In the quantum language, this new state would have the same ``eigennumber'' as the original. Any finite turn-on time $\tau$ will cause transitions away from this transport. Our goal is to minimize these transitions for all $\tau$. In passing let us note that for single-particle systems an alternative approximate approach based on classical flow field was recently developed in Refs.~\cite{Patra_2017, Patra_2021}.

\subsection*{Ansatz}

First, we must choose an ansatz for the form of our driving term. Following Ref.~\cite{Claeys_2019}, we can take a series expansion of the AGP, truncate the series at a finite order, and then insert this into our variational procedure. In the classical limit, this ansatz takes the form
\begin{align}
	\mathcal{A}_{\beta}^{(\ell)}= \sum_{k=1}^{\ell} (-1)^{k} \gamma_{k} \underbrace{\{H_0,\{H_0, \ldots\{H_0}_{2 k-1}, \partial_{\beta} H_0\}\}\} .
	\label{PB expansion}
\end{align}
where $\ell$ is the order at which we truncate the expansion, $\gamma_k$ are the variational parameters that we will optimize as functions of $\beta$, and the factor $(-1)^k$ is introduced for convenience. Here we will only consider ansatzes with one and two driving parameters, so we have
\begin{align}
	\mathcal{A}_{\beta}^{(1)} &= \gamma_1 x^3 p , \nonumber \\
	\mathcal{A}_{\beta}^{(2)} &= \gamma_1 x^3 p + \gamma_2 \left( -6 p^3 x + 10 p x^3 + 12 \beta p x^5 \right) .
	\label{nlo comm ansatz}
\end{align}
 Notice that the last term in the expression for $\mathcal{A}_{\beta}^{(2)}$ is proportional to $\beta$ and vanishes in the limit $\beta\to 0$. Therefore it is not very important at least at small values $\beta$, and we will exclude it from the variational ansatz. The $p^3 x$ term in $\mathcal{A}_{\beta}^{(2)}$ is identical to $\mathcal{A}_{\beta}^{(2)}$ so they can be merged together. Therefore for the remainder of the paper we will refer to our first- and second-order ansatzes as
\begin{align}
	\mathcal{A}_1^* &\equiv \tilde \gamma_1 x^3 p , \nonumber \\
	\mathcal{A}_2^* &\equiv \tilde\gamma_1 x^3 p +\tilde \gamma_2 x p^3,
	\label{nlo agps}
\end{align}
where $\tilde\gamma_1$ and $\tilde\gamma_2$ are our new variational parameters.

Let us point out that there is an alternative way to justify this variational ansatz by finding the AGP in the harmonic limit $\beta\to 0$. In Refs.~\cite{Jarzynski_CM_chaos, Sugiura_2020} it is shown that AGP can be represented through the following time integral:
\begin{align}
	\mathcal{A}_{\beta}=-\lim _{\epsilon \rightarrow 0^{+}} \frac{1}{2} \int_{-\infty}^{\infty} d t \operatorname{sgn}(t) \mathrm{e}^{-\epsilon|t|}\left(\partial_{\beta} H_0\right)(t) ,
	\label{integral rep}
\end{align}
where in quantum language $\left(\partial_{\beta} H_0\right)(t)$ is the Heisenberg representation of the operator $\partial_{\beta} H_0$, which in the classical language translates to the function $\partial_{\beta} H_0$ evaluated on time-dependent trajectories, which are solutions to the equations of motion (EoM) $(x(t),p(t))$ for the full nonlinear Hamiltonian. In the leading order in perturbation theory we can use the solutions of the non-interacting problem $\beta=0$:
\begin{align}
	x(t) = x \cos t + p \sin t , \nonumber \\
	p(t) = p \cos t - x \sin t ,
\end{align}
where $x$ and $p$ correspond to time-independent Schr\"odinger operators in the quantum language. In this limit Eq.~\eqref{integral rep} yields the following form for the AGP:
\begin{align}
	\mathcal{A}_{\beta} \rightarrow - \frac{5}{32} x^3 p - \frac{3}{32} x p^3 .
	\label{nlo integral nums}
\end{align}
We thus see that the exact AGP at $\beta\to 0$ is described by the ansatz~\eqref{nlo agps} with a particular choice of the coefficients $\tilde \gamma_1$ and $\tilde \gamma_2$. At $\beta>0$ this ansatz is no longer exact, but we can still variationally optimize $\tilde \gamma_1$ and $\tilde \gamma_2$ hoping that the resulting AGP will accurately describe the nonlinear system. We checked that both ansatzes~\eqref{nlo comm ansatz} and~\eqref{nlo agps} lead to a similar performance, so the extra $px^5$ appearing in Eq.~\eqref{nlo comm ansatz} has a very small effect on the ACD protocol.

\subsection*{Optimization}

As we discussed above, in order to perform optimization of the variational parameters, we slowly evolve the initial distribution corresponding to a microcanonical ensemble at $\beta=0$ to a finite value of nonlinearity and then minimize the action~\eqref{action def}, sampling from this probability distribution. In quantum language, this procedure corresponds to sampling from an instantaneous energy state adiabatically connected to the initial state of the harmonic oscillator. We discuss details of this procedure and show the obtained functional forms of the variational coefficients $\gamma_i(\beta)$ in Appendix \ref{app:num opt}.

\subsection*{Figures of Merit}

Once we simulate the driven system, we need a method for evaluating the performance of the given ansatz. In quantum systems, the golden standard is the fidelity of the final state with the target state. This figure of merit (FoM) tells us exactly how close we are to achieving adiabatic transport, as it not only ensures we approach a stationary state but also ensures that it is the \textit{correct} stationary state. In the classical regime, this task becomes much harder. In principle, we could evaluate some distance such as the  KL-divergence between the obtained and target probability distributions. But finding this distance is not easy, especially when we extend this analysis to systems with many degrees of freedom. We find that there is a much simpler measure, which is the energy variance of the final Hamiltonian in the obtained distribution. It works both in quantum and classical setups and is easy to evaluate. If we start in an exact microcanonical state and evolve it adiabatically, we should also end up in a microcanonical state such that the energy variance is zero. Similarly, in quantum mechanics if we start from a single eigenstate and adiabatically evolve in time we should remain in a single eigenstate and the energy variance is again zero. In both cases, non-zero energy variance indicates that the evolution was not adiabatic. For the remainder of this paper, we will focus on this FoM. We also checked other FoMs like fidelity or temporal fluctuations of observables and found that they contain similar information (see Appendix \ref{app:FoM} for more details.)

\subsection*{Results}

\begin{figure}
	\centering
	\includegraphics[width=8.6cm]{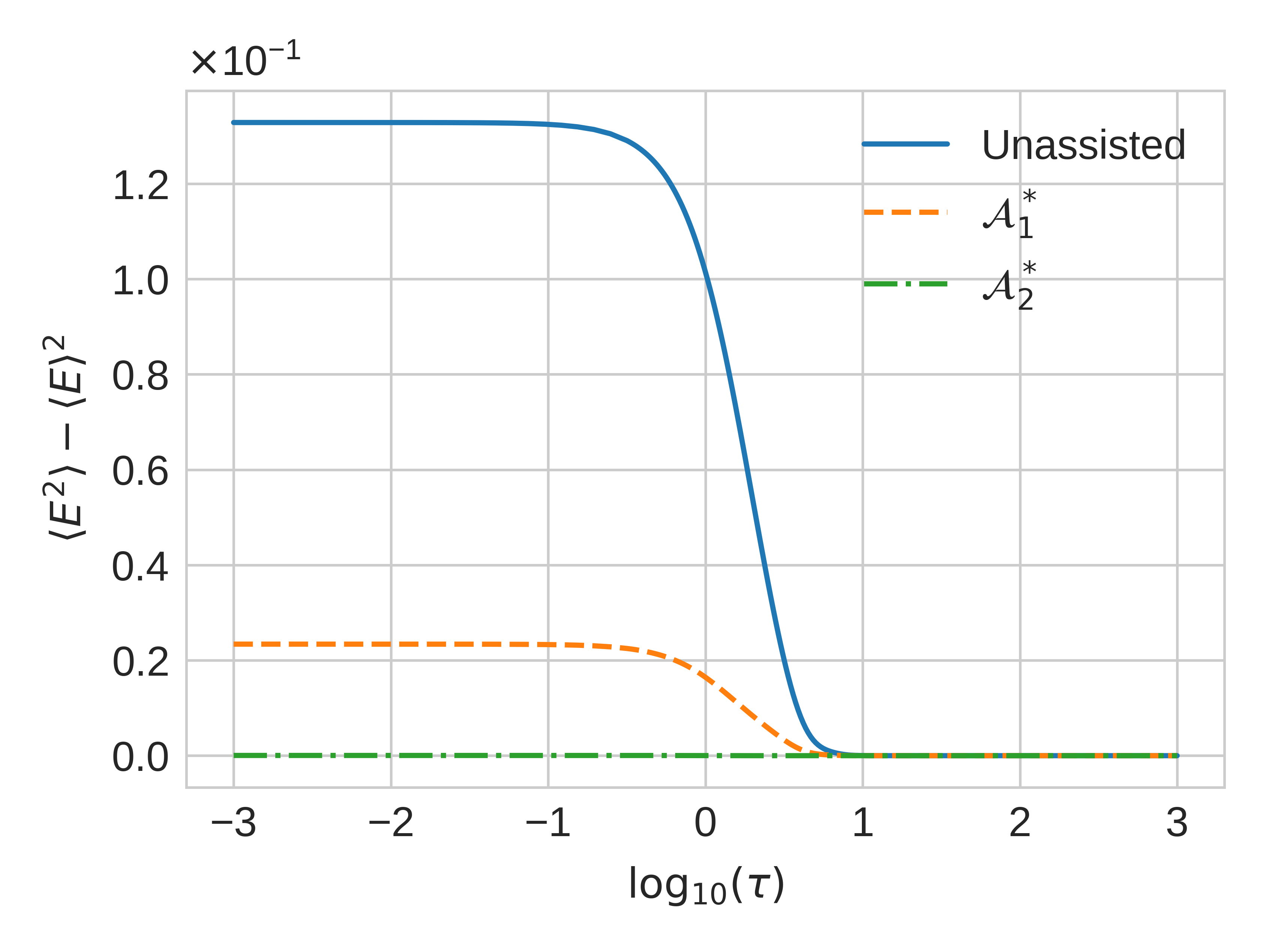}
	\caption{Final energy variance after ramps of $\beta$ of various durations $\tau$ for a single anharmonic oscillator. Orange and green lines correspond to first- (second-) order CD driving protocols. In the limit of $\tau\to 0$, which corresponds to driving only with the AGP, the first-order CD protocol gives the energy variance suppression of the order of $5.7$ compared to the unassisted (quench) protocol; the second-order CD protocol increases this suppression to a factor of $\sim 2700$.}
	\label{fig:cnlo statevar}
\end{figure}

In Fig. \ref{fig:cnlo statevar}, we show the results of implementing the ACD protocols with the optimized AGPs and compare their performance against a bare, unassisted ramp for various values of the turn-on time $\tau$. In these simulations, we initialize our system in a microcanonical distribution of $E_0 = 1$ and evolve while $\beta$ grows according to the protocol \eqref{beta protocol}.

Note that, as expected, all turn-on curves interpolate between some asymptotic value at small $\tau$ and zero at large $\tau$. All protocols in the limit $\tau\gg 1$ give zero energy variance in agreement with the adiabatic theorem. In the opposite $\tau\to 0$ limit, which corresponds to the instantaneous (quenched) protocol for unassisted driving and the protocols where one drives the system only with the AGP for ACD driving, we see the largest difference between the three protocols. In particular, in this limit the unassisted protocol gives the largest energy variance, and the first-order ACD suppresses the energy variance by an approximate factor of $5.7$, which pales in comparison to the performance of the second-order ACD, which suppresses the energy variance by a factor of $2700$. We thus see that the addition of the rather unusual $x p^3$ to the Hamiltonian yields a dramatic improvement in its performance. The difference becomes even larger if we analyze the quantum fidelity of the state preparation or the temporal variance of the harmonic energy as FoMs (see Appendix~\ref{app:FoM}). Incidentally, this $xp^3$ term also appears in the exact CD driving of a potential in the shape of a KdV soliton moving in a gas of non-interacting quantum particles~\cite{Okuyama_2017, Kolodrubetz_2017}. Clearly, the $xp^3$-term plays a crucial role in creating the efficient CD driving protocol. However, this presents a unique challenge when implementing this protocol in experiments, as such terms do not naturally appear in typical Hamiltonians. These terms can be, however, designed using periodic lattice potentials, where the dispersion relation generally contains $\sin(p)$ type terms, where $p$ is the lattice momentum. 

\paragraph*{Floquet Implementation.}
Generally the expansion~\eqref{PB expansion} can be implemented through the experimentally accessible Floquet protocols~\cite{Claeys_2019}. Such protocols only involve $H_0$ and $\partial_\beta H_0$ as driving terms. Specifically choosing 
\begin{align}
	H_{FE}(t) = f(t) H_0 + g(t) \partial_{\beta} H_0,
\end{align}
where $f(t)$ and $g(t)$ are periodic functions in time, one can always realize a CD protocol with the AGP given by Eq.~\eqref{PB expansion} with arbitrary time-dependent coefficients $\gamma_k$. 
If the function $f(t)$ is positive at any time then one can further simplify the protocol by rescaling time. Namely, dividing the Schr\"odinger equation $i\hbar \partial_t \psi = H_{FE}(t) \psi$
by $f(t)$. One can define a new ``physical'' time as  
\begin{equation}
\label{eq:tau_t}
    \tau(t) = \int_{0}^{t} f(t') dt'.
\end{equation}
Then
the Schr\"odinger equation becomes
\begin{equation}
i \hbar \partial_\tau \psi=\tilde H_{FE}
(\tau)\psi,\quad	\tilde{H}_{FE}(\tau) = H_0 + \kappa(\tau) \partial_{\beta} H_0 ,
\label{eq:Floquet_CD_kappa}
\end{equation}
where $t$ is now a parameter, which depends on the physical protocol time $\tau$ through inverting Eq.~\eqref{eq:tau_t} and $\kappa(\tau) = g(t(\tau))/f(t(\tau))$. Thus, we can implement the Floquet protocol by only varying the coupling $\partial_{\beta} H_0$ in time. 

Note that generally the Floquet protocols found in Ref.~\cite{Claeys_2019} require non-positive $f(t)$ and the mapping~\eqref{eq:tau_t} is not possible without inverting the sign of $H_0$. Finding a class of general protocols with $f(t)>0$ remains an open problem. In Ref.~\cite{Villazon2019Swift} it was solved for a particular setup leading to a finite speed limit.


\section{$\beta$-FPUT CHAIN}

\label{sec:bfput}

\begin{figure}
	\centering
	\includegraphics[width=8.6cm]{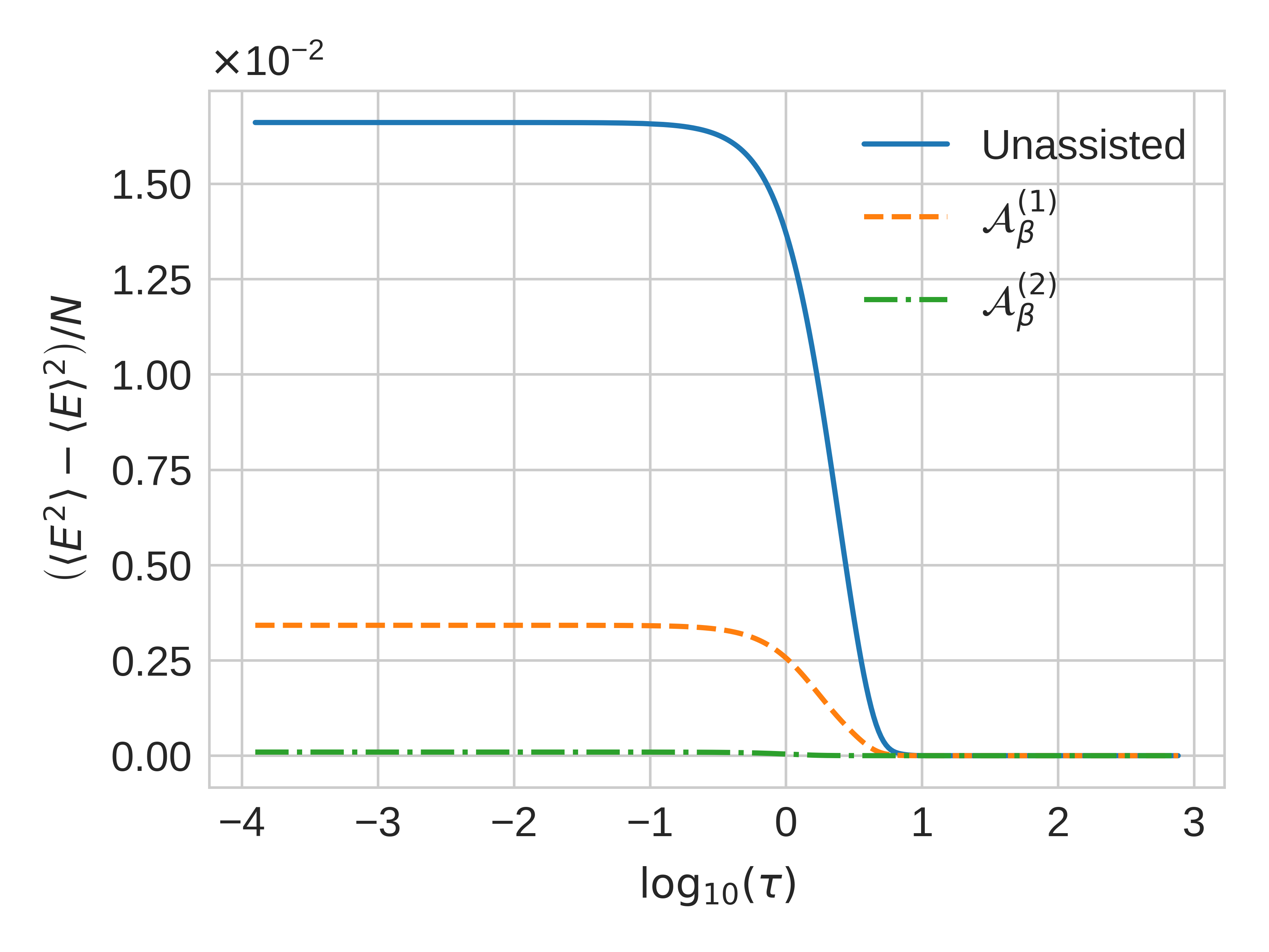}
	\caption{Final energy variance (over system size) after turn-ons of various durations $\tau$ for the $\beta$-FPUT chain with $N=2$ active sites. $\mathcal{A}_{\beta}^{(1)}$ corresponds to first-order driving using the Poisson bracket expansion and suppresses energy variance by a factor of $\sim 4.9$ relative to no driving. $\mathcal{A}_{\beta}^{(2)}$ corresponds to second-order driving and suppresses energy variance by a factor of $\sim 176$ relative to no driving.}
	\label{fig:bfput2 statevar}
\end{figure}

Now that we have a method for implementing ACD driving in classical oscillators, we want to apply it to the more complex $\beta$-FPUT chain. The Hamiltonian for this system is given by
\begin{align}
	H_0 = \sum_{n = 1}^{N} \frac{p_n^2}{2} + \sum_{n=0}^{N} \frac{1}{2} (q_{n+1} - q_n)^2 + \frac{\beta}{4} (q_{n+1} - q_n)^4 ,
	\label{fput ham}
\end{align}
where $N$ is the number of active oscillators in the system. We work with fixed boundary conditions, such that $p_0 = p_{N+1} = 0$ and $q_0 = q_{N+1} = 0$. We initialize our state in a single normal mode with a fixed momentum $k$ keeping all other modes empty. The normal modes are defined through a canonical transformation
\begin{align}
	q_n = \sqrt{\frac{2}{N+1}} \sum_{k=1}^{N} Q_k \sin \left( \frac{n k \pi}{N+1} \right) ,
	\label{mode transform}
\end{align}
and similarly for $p_n$ and $P_k$. As long as $\beta = 0$, each of these modes corresponds to a stationary orbit in phase space with a conserved energy $E_0^{(k)}= P_k^2/2+\omega_k^2 Q_k^2/2$, where $\omega_k=2\sin(\frac{\pi k}{2 (N+1)} )$. Before proceeding to a general case, we will analyze a simpler two oscillator system with $N=2$, which is already chaotic in the presence of nonlinearity and hence does not have a closed form AGP. We will choose a normal mode with $k=1$ initially populated with the energy $E_0^{(k)}\equiv E_0=1$.

If we compute the AGP in the noninteracting point we will get four independent terms. This number grows rapidly if we add more oscillators. Keeping all of these terms in the variational ansatz is costly. For this reason we will use the full ansatz~\eqref{PB expansion} truncating the expansion at the second-order and optimizing for the two variational coefficients $\gamma_1$ and $\gamma_2$.

Following the same process outlined in Section \ref{sec:nlo}, we optimize and simulate our ACD driving in the $N=2$ FPUT chain. We plot the results in Fig. \ref{fig:bfput2 statevar}. We can see that essentially the same results are produced: first-order driving suppresses energy variance over the unassisted protocol by about a factor of $5$ in the $\tau\to 0$ limit. The second-order driving gives a significantly more substantial improvement, reducing the energy variance by a factor of $\sim 180$. This improvement is somewhat less than the one for the single-oscillator case, where the suppression factor was $\sim 2700$, but it is still fairly impressive and almost completely eliminates non-adiabatic transitions.

\begin{figure}
	\centering
	\includegraphics[width=8.6cm]{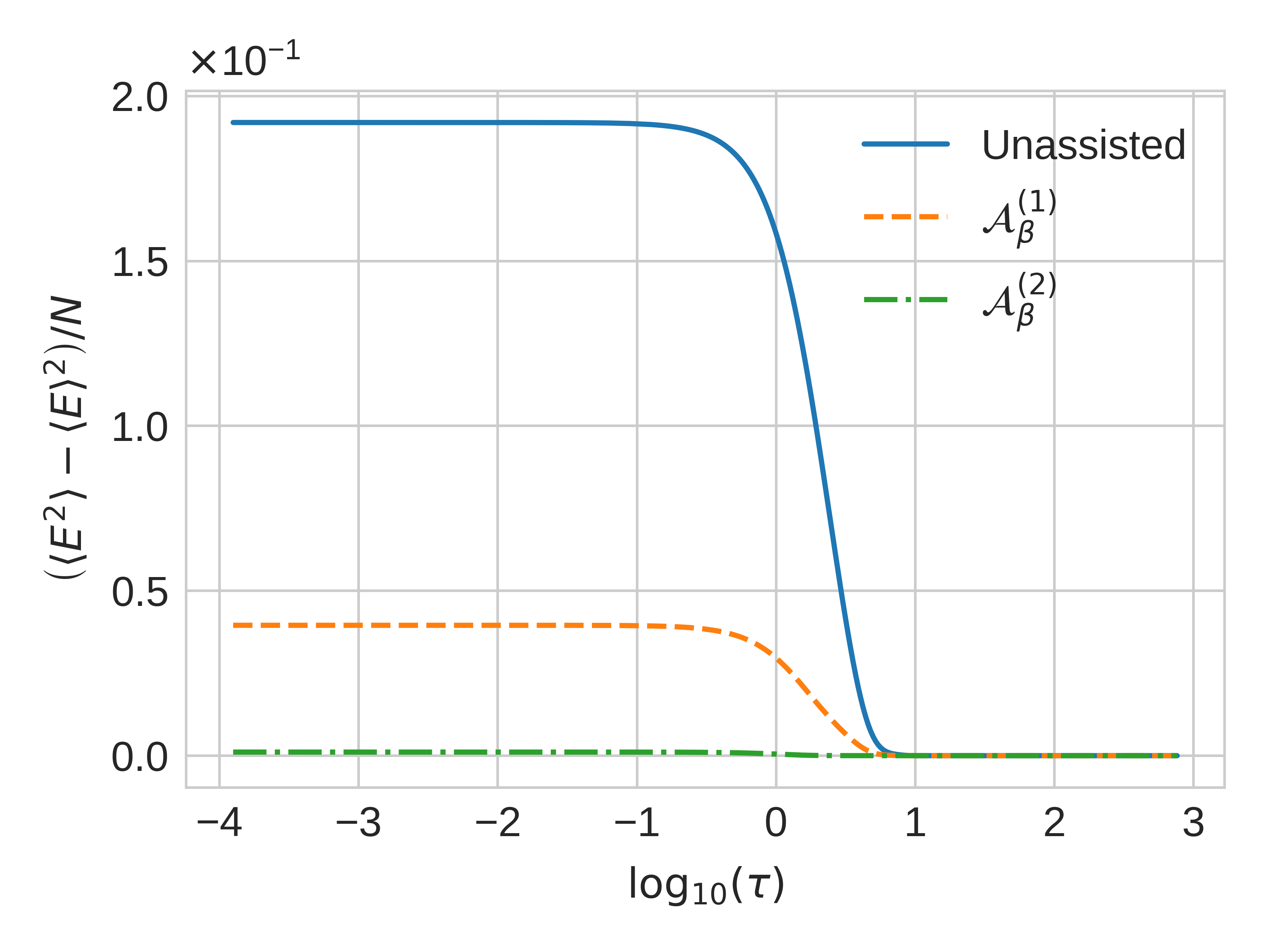}
	\caption{Final energy variance (over system size) after turn-ons of various durations $\tau$ for the $\beta$-FPUT chain with $N=50$ active sites. $\mathcal{A}_{\beta}^{(i)}$ still correspond to $i$th order driving using the Poisson bracket expansion and have the same factors of suppression as Fig. \ref{fig:bfput2 statevar}.}
	\label{fig:bfput50 statevar}
\end{figure}

We can now extend the analysis to the thermodynamic limit $N\to \infty$. This limit is well defined if we choose initial conditions with a fixed wavelength of the excited mode and the fixed amplitude of the normal mode in the initial state, which also corresponds to a fixed momentum and energy densities in the system. Specifically we choose the initial state where the initial momenta of particles are zero ($p_n=0$) and the initial displacements of particles are given by $q_n = A \sin (n k \pi / (N+1))$. A stationary probability distribution, where we want to initialize the system and which corresponds to a quantum stationary state, is obtained by time-averaging the initial probability distribution with the initial non-interacting Hamiltonian $H_0(\beta=0)$. In practice this means that before starting the ramp of the nonlinearity one has to evolve the system with $H_0(\beta=0)$ for a random time within one period of oscillation $T_k=2\pi/\omega_k$ (for our simulations, we implemented this by choosing a set of times uniformly distributed between 0 and $T_k$ and averaging over them). It is easy to see that the wavelength and the energy corresponding to these states are given by 
\begin{align}
	\lambda &= \frac{2 (N+1)}{k} , \nonumber \\
	E_0 &= A^2 (N+1) \sin^2 \left( \frac{k \pi}{2(N+1)} \right) .
\label{eq:lambda_E0_FPUT}
\end{align}
Now consider our two-site problem. Plugging in our values of $N=2$, $E_0 = 1$, and $k = 1$, we see that our wavelength and amplitude are $\lambda = 6$ and $A = 2/\sqrt{3}$. To approach our desired thermodynamic limit, we have to find a way to increase $N$ while keeping $\lambda$ and $A$ fixed. This can be done by fixing the amplitude at $A=2/\sqrt{3}$ and the wavelength at $\lambda=6$ and choosing $N=k\lambda/2-1\;\leftrightarrow \; k=2(N+1)/\lambda$, where $k$ is an integer. It is easy to see that this choice corresponds to the system size-independent energy density in the system $E_0=k\propto N$. It is expected that the protocol performance quickly becomes $N$-independent as we are dealing with a local Hamiltonian, use the local ansatz for the AGP and local initial conditions. The only way the system knows about its size is through the boundary conditions, which are expected to play a negligible role at large $N$. For concrete calculations we choose $k=17$, corresponding to $N=50$. We checked that the results indeed do not change if we further increase $N$ and $k$. We show the results of such simulations in Fig. \ref{fig:bfput50 statevar}. Aside from the different scale on the vertical axis, they look identical to Fig. \ref{fig:bfput2 statevar}. The relative suppression between each curve is also preserved: increasing $N$ does not sacrifice the performance of the ACD driving. We also note that the optimized driving coefficients $\gamma_i(\beta)$ show essentially no $N$-dependence, testifying to the stability and applicability of the method.


\section{LONG WAVELENGTH INSTABILITY}
\label{sec:instability}

\begin{figure}
	\centering
	\includegraphics[width=8.6cm]{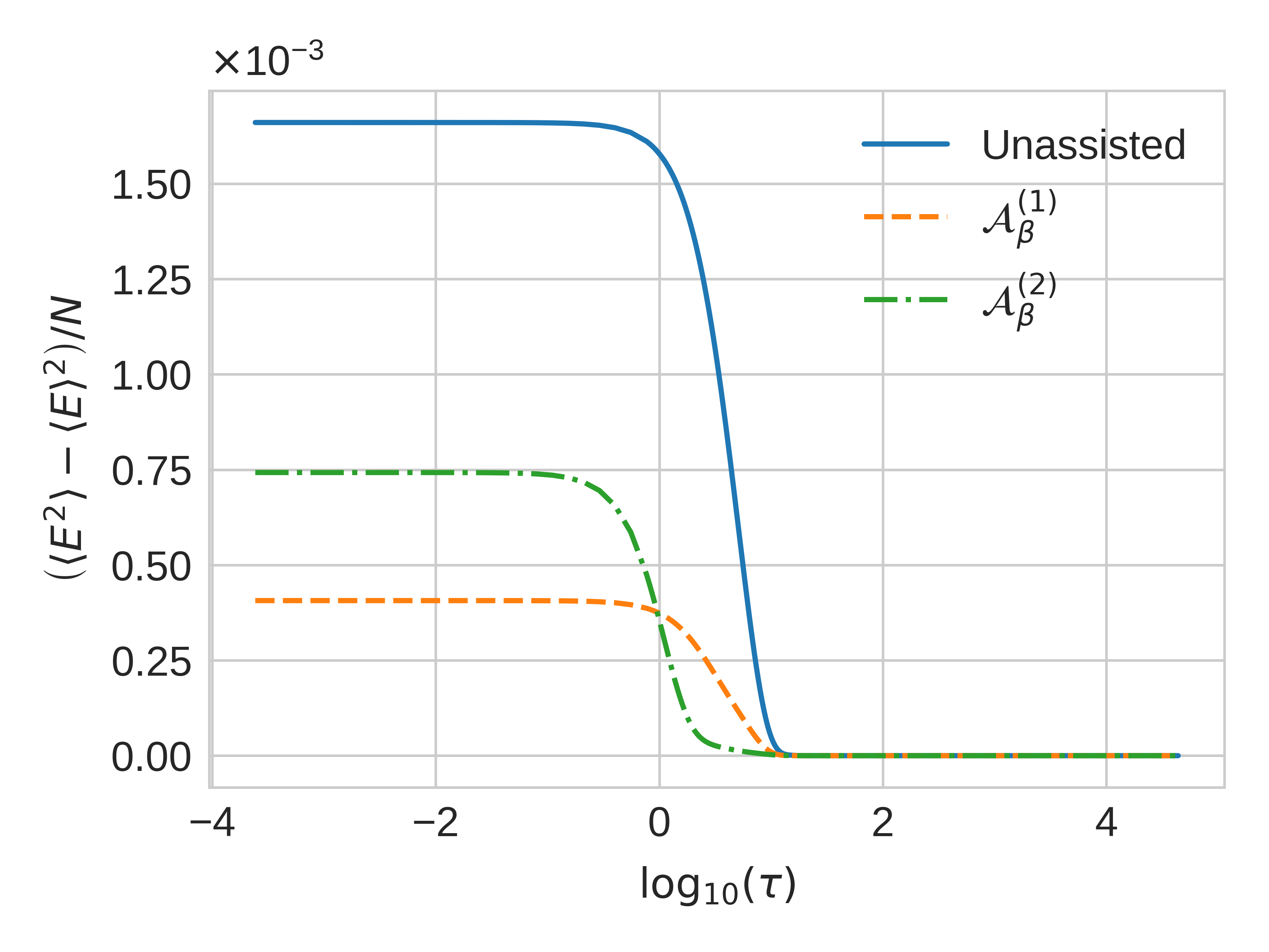}
	\caption{Final energy variance (over system size) after turn-ons of various durations $\tau$ for the $\beta$-FPUT chain with $N=5$ active sites. Rather than hold wavelength and amplitude fixed as $N$ increases, we keep all our initial mode number and energy fixed at $1$. We clearly run into an instability at second-order driving in this limit.}
	\label{fig:bfput5 statevar}
\end{figure}

\begin{figure}
	\centering
	\includegraphics[width=8.6cm]{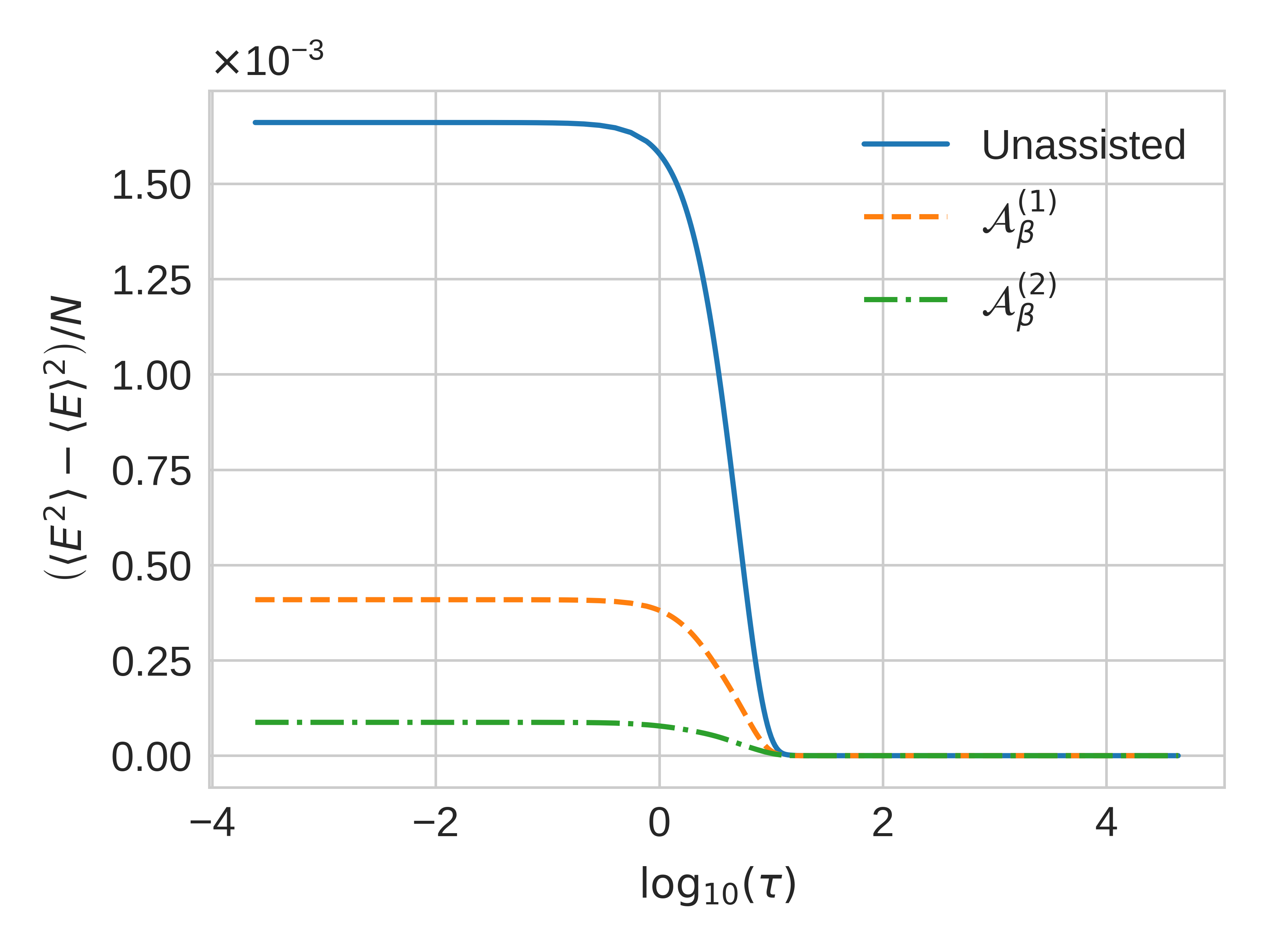}
	\caption{Final energy variance (over system size) after turn-ons of various durations $\tau$ for the $\beta$-FPUT chain with $N=5$ active sites. We fight the instability in Fig. \ref{fig:bfput5 statevar} by evaluating the action along a distribution of adiabatic trajectories over a broadened distribution of initial energies (see text for details).}
	\label{fig:bfput5 width statevar}
\end{figure}

Traditionally, the FPUT problem was analyzed for a very special initial condition corresponding excitation only in the first mode, $k=1$. Note that as we fix the total energy in the system given by Eq.~\eqref{eq:lambda_E0_FPUT}, the amplitude $A$ scales as $\sqrt{N}$ as one increases the system size. Such a state is not very stable thermodynamically. This instability is precisely reflected in the very large amplitude oscillations in the mode occupancy, which emerge after quenching the nonlinearity ~\cite{fput_original,Tuck_Menzel,Pace_Reiss_2019}. In other words, studying the FPUT problem is equivalent to turning on $\beta$ instantaneously in our setup. Note that should we fix the energy density while keeping $k=1$, $A$ would scale even faster, linearly with $N$, leading to even stronger instability. For $N=2$ this initial condition is identical to the one we used earlier, but as $N$ increases we start to see a difference. In particular, we find that the second-order driving performs slightly worse and worse in terms of suppression of the energy variance until we reach $N = 5$, where the variational protocol essentially breaks down as illustrated in Fig. \ref{fig:bfput5 statevar}. Our optimization method, which previously led to dramatic suppression of excitations, has now gone awry and made our second-order ansatz perform \textit{worse} than the first-order ansatz in the quench limit $\tau\to 0$. Note however that at intermediate speeds $\tau\geq 2$, the second-order ACD protocol strongly outperforms the first-order one, which indicates that some instability develops as the protocol duration gets shorter.


This failure of the second-order ACD protocol looks somewhat mysterious at first sight, so we checked and confirmed that it is not due to some artifact of the numerical integration. We found that the origin of the failure of the variational method in this case is high sensitivity of the dynamical response of the system to the energy of the state. In particular, the ACD protocol tries to keep the system close to the adiabatically connected state. However, because the variational AGP does this only approximately, the actual state of the system under fast ACD protocol deviates from the adiabatically connected state. Normally this deviation is not important, as the variational AGP is very stable with respect to small changes in the system. But for our thermodynamically unstable state, this is not the case, and the evolution is highly sensitive to the precise form of the AGP, so any deviation of the probability distribution from the target one is amplified. We found that even in this situation one can fix the ACD protocol by choosing the initial distribution with fluctuating energy. Specifically we choose an initial Gaussian ensemble, where we populated the first $k=1$ mode with the mean energy $\bar E_0=1$ and the standard deviation $d_E=0.4$. As the energy must be always positive we suppressed the tail of the distribution corresponding to $E_0<0$. The total action is then defined as the weighted average of the actions corresponding to different values of $E_0$ weighted with this distribution. For calculations we used a set of 40 different realizations of $E_0$. As illustrated in Fig.~ \ref{fig:bfput5 width statevar}, the outlined procedure stabilizes performance of the ACD driving, although the enhancement of the second-order AGP is not as pronounced as in the previous thermodynamically stable case of initial constant $\lambda$ and energy density. These results suggest that in nontrivial situations where ACD is used to drive the system through some unstable regime, one needs to carefully choose the distribution $P_\beta(\vec q,\vec p)$ defining the action: too broad a distribution might lead to poor performance of the ACD protocol, while too narrow a distribution might lead to developing instabilities at short protocols times. In Appendix \ref{app:sensitivity}
we provide a more detailed analysis of the origin of this instability.

%

\section{DISCUSSION AND CONCLUSIONS}

Counter-diabatic driving can find applications in a broad range of systems as a way to control the evolution of a state under changes to the Hamiltonian. With ACD driving, one can find protocols suppressing excitations even when the protocol durations are short and unassisted protocols lead to large irreversible losses. In this work we have successfully generalized ACD driving from quantum spin systems to classical nonlinear oscillator systems. In particular, we have shown that ACD driving can dramatically suppress transitions in both a single anharmonic oscillator and the $\beta$-FPUT chain for the cost of a couple driving parameters. We showed that our results survive in the thermodynamic limit and do not require fine-tuning for generic initial states corresponding to populating a normal mode with fixed system size-independent wavelength and amplitude.

We found that the dramatic suppression of dissipation at short protocol durations requires adding counterterms to the Hamiltonian, which are cubic in momentum. Such terms are rather unusual and do not naturally appear in standard setups. However, they can be engineered either in periodic potentials having non-quadratic dispersion or using Floquet protocols. Precise implementation of such ACD protocols might depend on the details of the system and available controls. Our work shows that ACD can be potentially extended to both quantum and classical systems of interacting particles and fields. Because the AGP can be also used to find effective degrees of freedom~\cite{Wurtz_SW_2020}, our work shows considerable potential for finding effective non-perturbative low energy descriptions of such systems as well as efficient computational methods of simulations of these systems.

\section*{ACKNOWLEDGEMENTS}

We would like to thank Kevin A. Reiss, Salvatore D. Pace, Andreas Hartmann, and Dries Sels for stimulating discussions. We also thank Boston University's Research Computing Services for their numerical resources. N.O.G. is grateful for financial support from Boston University's Undergraduate Research Opportunities Program and Boston University's Physics Research Opportunity REU, funded by NSF Grant PHY-1852266. A. P. was supported by the NSF under Grants No. DMR-1813499 and DMR-2103658 and by the AFOSR under Grants No. FA9550-16-1-0334 and FA9550-21-1-0342.

\appendix
\section{Numerical Optimization}
\label{app:num opt}

Here we outline how we optimize $\gamma_i(\beta)$ for $\mathcal{A}_2^*$ in Eq. ~\eqref{nlo agps}. First, we can analytically expand $G_{\beta}$ into three terms:
\begin{align}
	G_{\beta} &= \partial_{\beta} H_0 - \gamma_1 \left\{ x^3 p , H_0 \right\} - \gamma_2 \left\{ x p^3 , H_0 \right\} \nonumber \\
	&= G_{00} + \gamma_1 G_{10} + \gamma_2 G_{01} .
	\label{G split terms}
\end{align}
When we square $G_{\beta}$ to find the action, we get
\begin{align}
	G_{\beta}^2 &= G_{00}^2 + \gamma_1^2 G_{10}^2 + \gamma_2^2 G_{01}^2 + 2 \gamma_1 G_{00} G_{10} \nonumber \\
	&+ 2 \gamma_2 G_{00} G_{01} + 2 \gamma_1 \gamma_2 G_{10} G_{01} .
	\label{G2}
\end{align}
It's also easy to see that
\begin{align}
	\langle G_{\beta}\rangle^2 &= \langle G_{00} \rangle^2 + \gamma_1^2 \langle G_{10} \rangle^2 + \gamma_2^2 \langle G_{01}\rangle^2 + 2 \gamma_1 \langle G_{00}\rangle \langle G_{10}\rangle \nonumber \\ 
	&+ 2 \gamma_2 \langle G_{00}\rangle \langle G_{01}\rangle + 2 \gamma_1 \gamma_2 \langle G_{10}\rangle \langle G_{01}\rangle .
	\label{G exp 2}
\end{align}
Taking the expectation value of Eq.~ \eqref{G2} and subtracting Eq. ~\eqref{G exp 2}, we get the action in Eq. ~\eqref{action def}. We can now write this object in the form
\begin{align}
	\mathcal{S}(\mathcal{A}_{2}^*) &= a_{00} + a_{10} \gamma_1 + a_{20} \gamma_1^2 \nonumber \\
	&+ a_{01} \gamma_2 + a_{02} \gamma_2^2 + a_{11} \gamma_1 \gamma_2 .
\end{align}
If we now adiabatically transport the system in our simulation, we can numerically compute $\{a_{ij}\}$ at each separate value of $\beta \in [0,1]$. For each of these values, we can independently optimize the action by requiring $\partial \mathcal{S} / \partial \gamma_i = 0$, which yields
\begin{align}
	\gamma_1 &= \frac{2 a_{02} a_{10} - a_{01} a_{11}}{a_{11}^2 - 4 a_{02} a_{20}} , \nonumber \\
	\gamma_2 &= \frac{2 a_{01} a_{20} - a_{10} a_{11}}{a_{11}^2 - 4 a_{02} a_{20}} .
\end{align}
At this point, our program will spit out an array of optimal values for each $\gamma_i$, with each element corresponding to the optimal value for a certain value of $\beta$ on our interval. We can now fit these arrays to functional forms, and we will have our optimal driving protocol. For the entirety of this paper, we used the fitting ansatz
\begin{align}
	\gamma(\beta) = \frac{b_0 + b_1 \beta + b_2 \beta^2 + b_3 \beta^3 + \cdots}{1 + c_1 \beta + c_2 \beta^2 + c_3 \beta^3 + \cdots} .
	\label{gamma ansatz}
\end{align}
We restrict $c_i \geq 0$ to improve the stability of the fit and bypass possible singularities. 

\begin{figure}
	\centering
	\includegraphics[width=8.6cm]{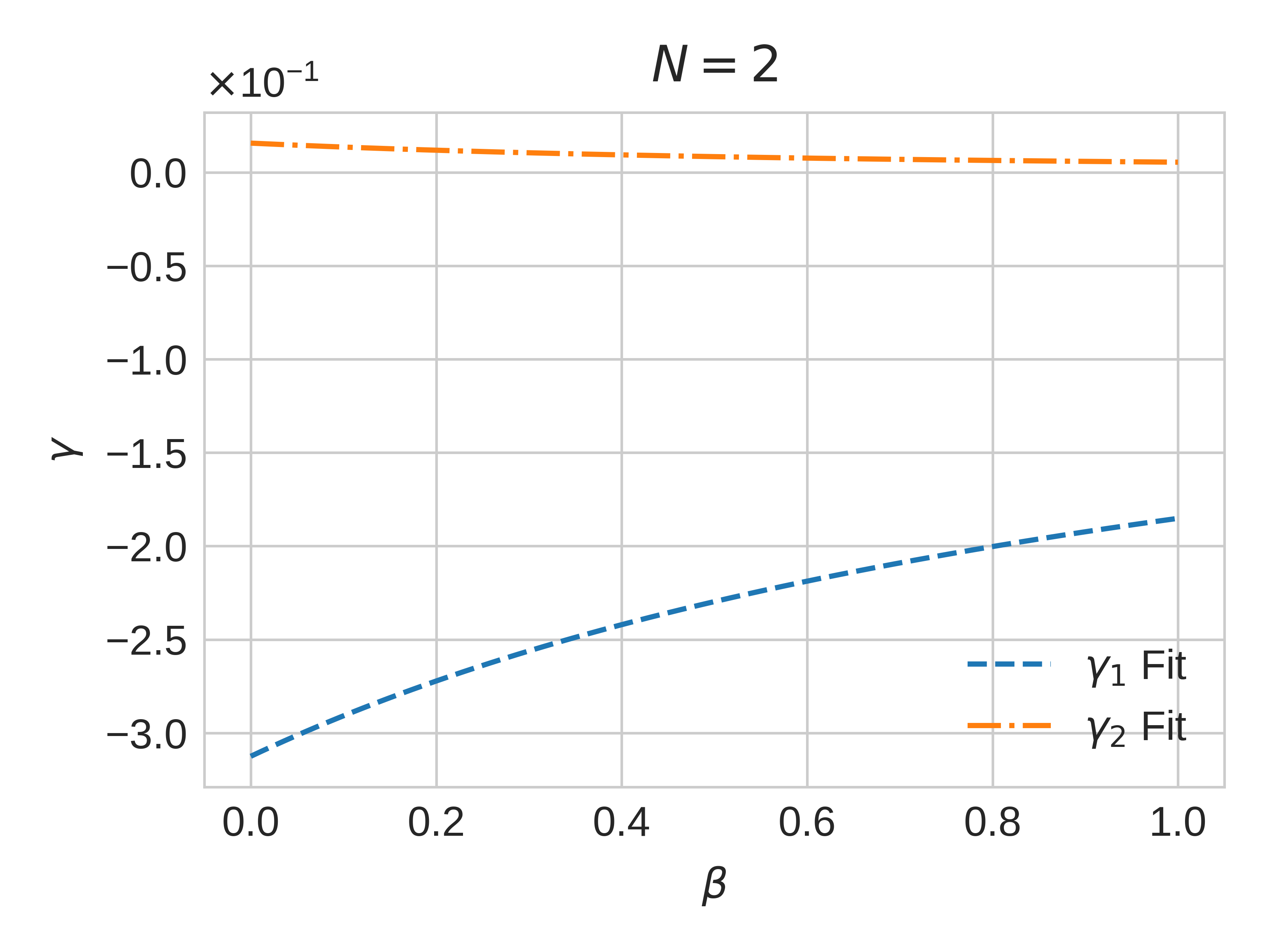}
	\caption{Driving coefficients $\gamma_i$ as functions of $\beta$ for the $N=2$ FPUT system discussed in the text. These are the functions spit out by the variational procedure after fitting the numerical data.}
	\label{fig:gvb N2}
\end{figure}

\begin{figure}
	\centering
	\includegraphics[width=8.6cm]{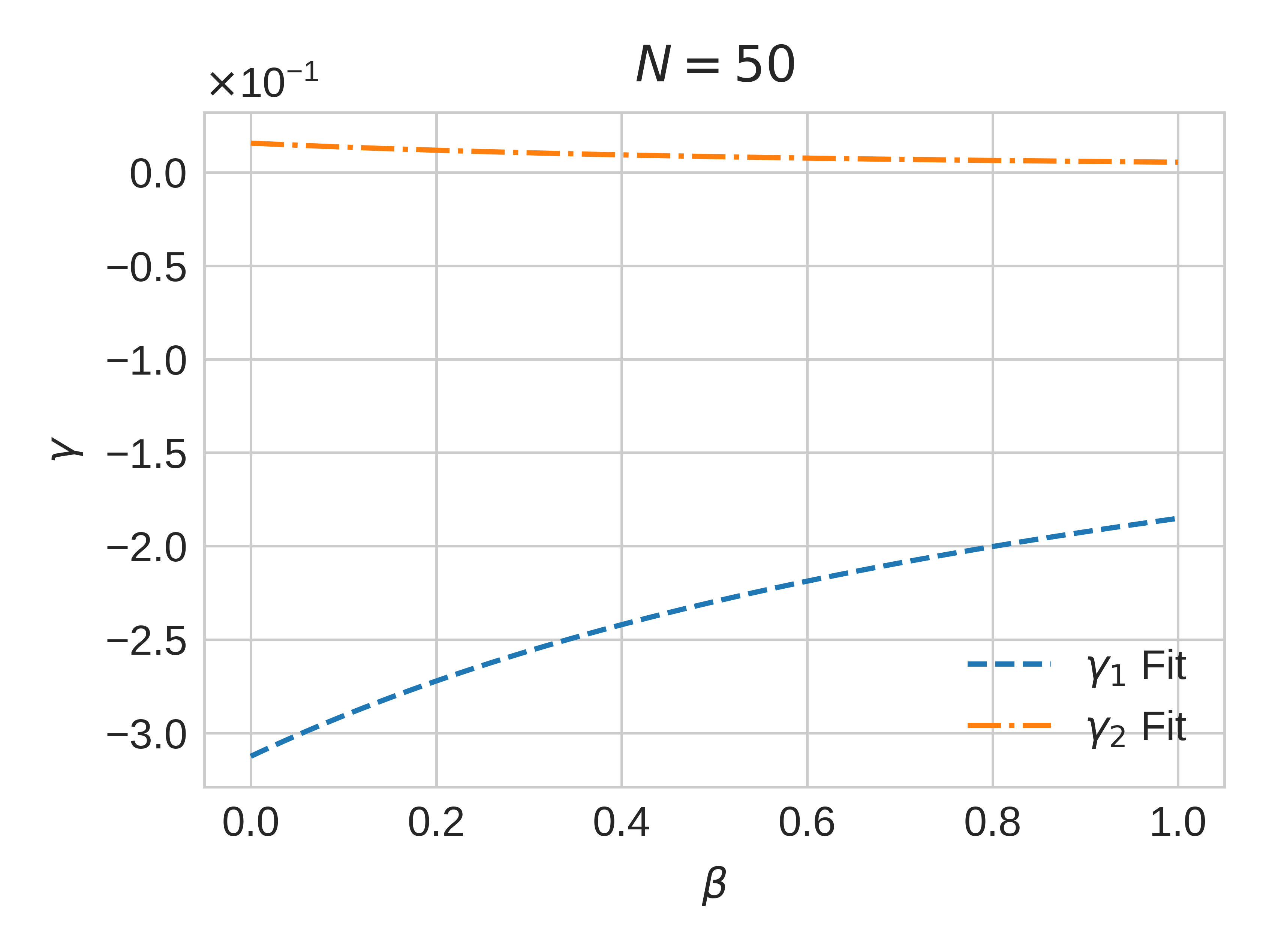}
	\caption{Driving coefficients $\gamma_i$ as functions of $\beta$ for the $N=50$ FPUT system discussed in the text. Note the essentially identical results compared to Fig. \ref{fig:gvb N2}.}
	\label{fig:gvb N50}
\end{figure}

In Fig. \ref{fig:gvb N2}, we plot the fitted driving coefficients $\gamma_i(\beta)$ as functions of the nonlinearity for the $N=2$ $\beta$-FPUT system discussed in Section \ref{sec:bfput}. Note that, although the magnitude of $\gamma_1$ is always significantly greater than $\gamma_2$, the presence of this second term allows us to achieve dramatic suppression with ACD driving. In Fig. \ref{fig:gvb N50}, we plot the same optimized coefficients but for the $N=50$ FPUT system discussed in the text. Note that these optimized functions are essentially identical to those in the $N=2$ case, indicating that our thermodynamic limit is stable and that the same driving parameters can be applied to systems with drastically different $N$.


\section{Figures of Merit}
\label{app:FoM}

Here we'll discuss other figures of merit we considered as ways to quantify non-adiabatic effects. To begin, we'll start in the quantum version of the single anharmonic oscillator. Here, we can compare our classical FoMs to the gold standard: the overlap of the final state with the target state. More specifically, we look into $1 - F^2$, where $F = \left| \bra{\psi_{\mathrm{target}}} \ket{\psi_f} \right|$ is the fidelity. Once this is calculated, we can compare it to $\bra{\psi_f} H_0^2 \ket{\psi_f} - \bra{\psi_f} H_0 \ket{\psi_f}^2$, the quantum version of the FoM used in the body of the paper, to determine whether or not the two measures give similar information about the performance of the ACD protocols.

\begin{figure}
	\centering
	
	\begin{subfigure}{\linewidth}
		\includegraphics[width=8.6cm]{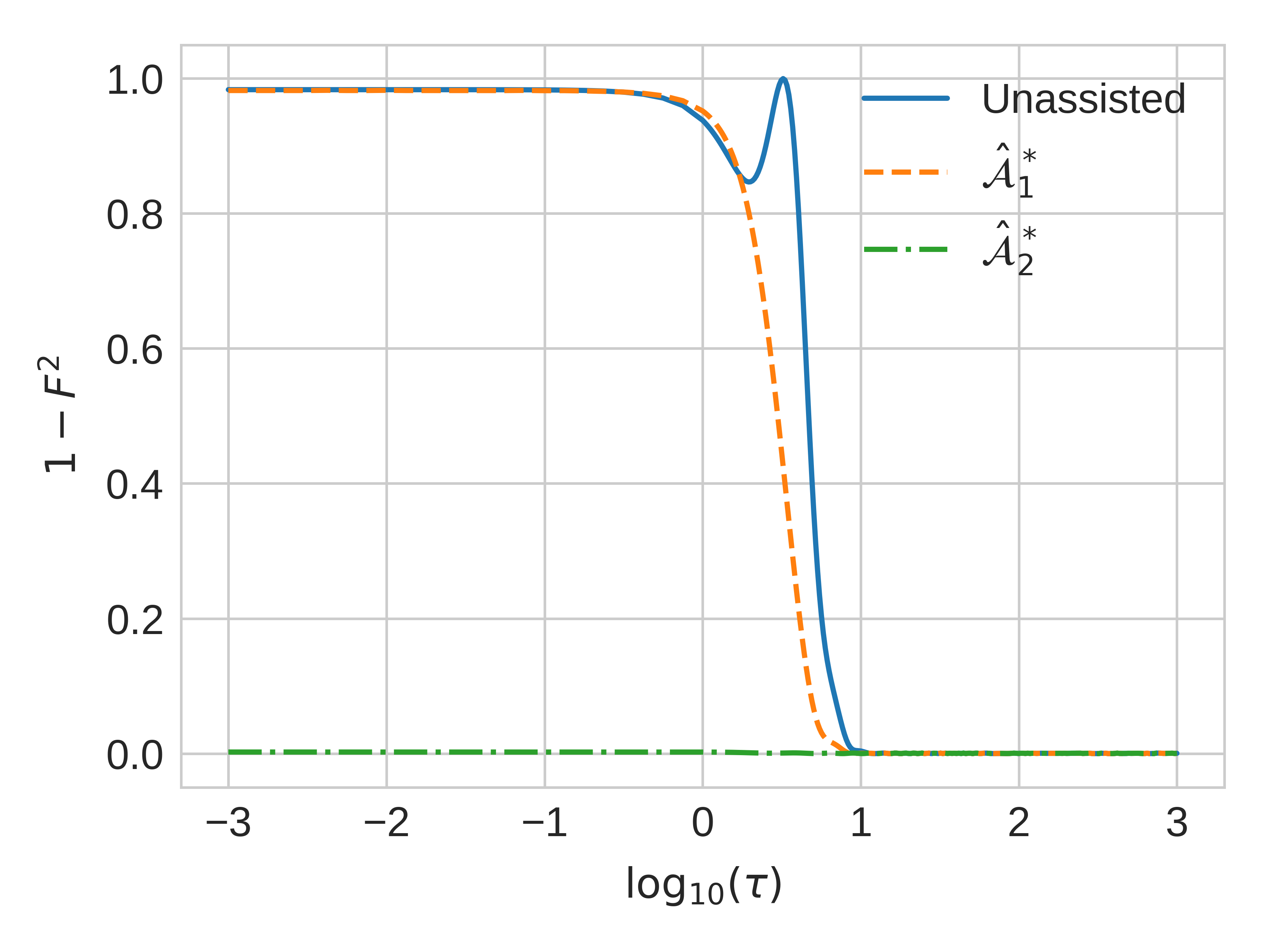}
		\caption{}
		\label{fig:qnlo minfid}
	\end{subfigure}
	
	\begin{subfigure}{\linewidth}
		\includegraphics[width=8.6cm]{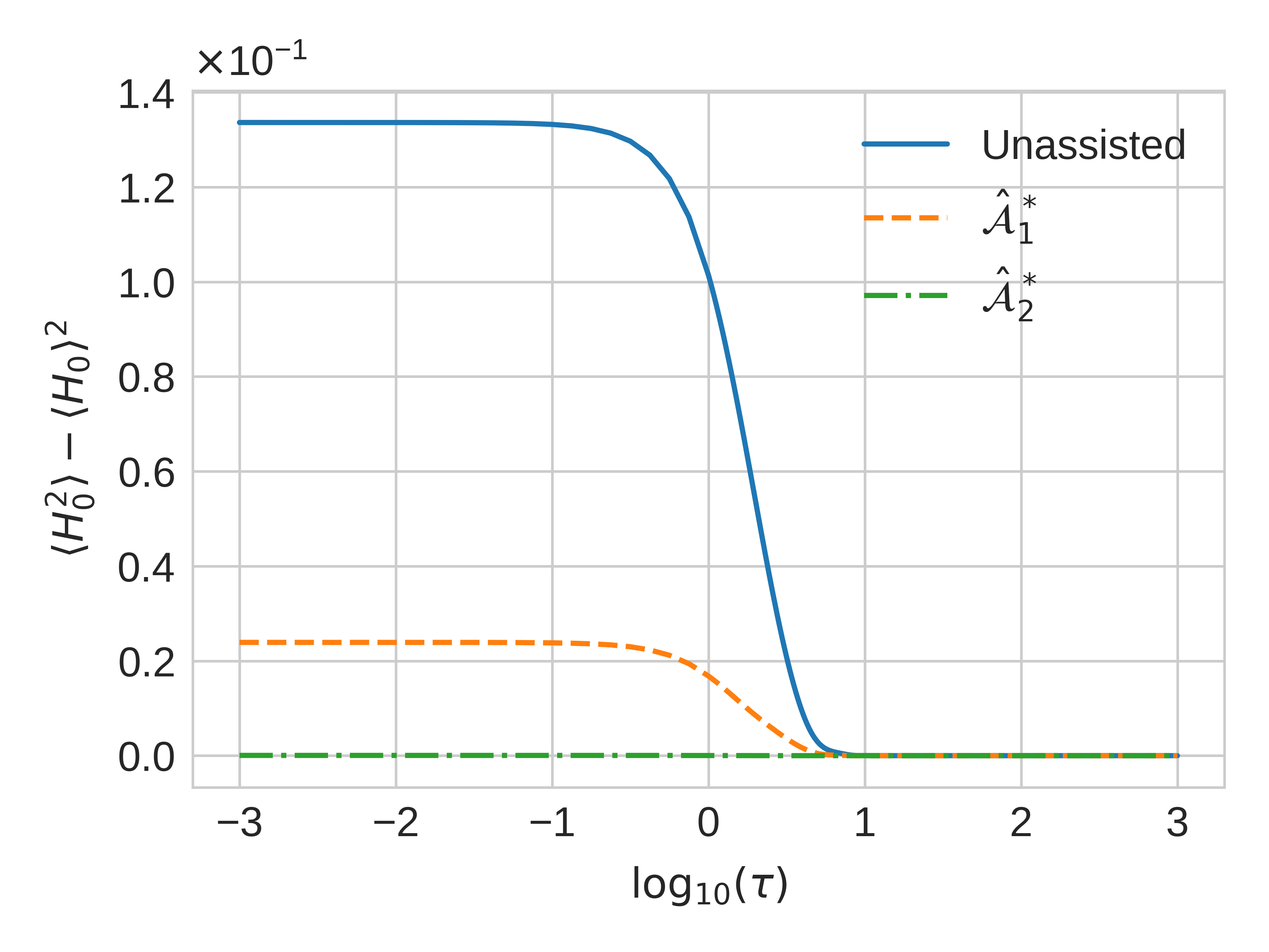}
		\caption{}
		\label{fig:qnlo statevar}
	\end{subfigure}
	
	\begin{subfigure}{\linewidth}
		\includegraphics[width=8.6cm]{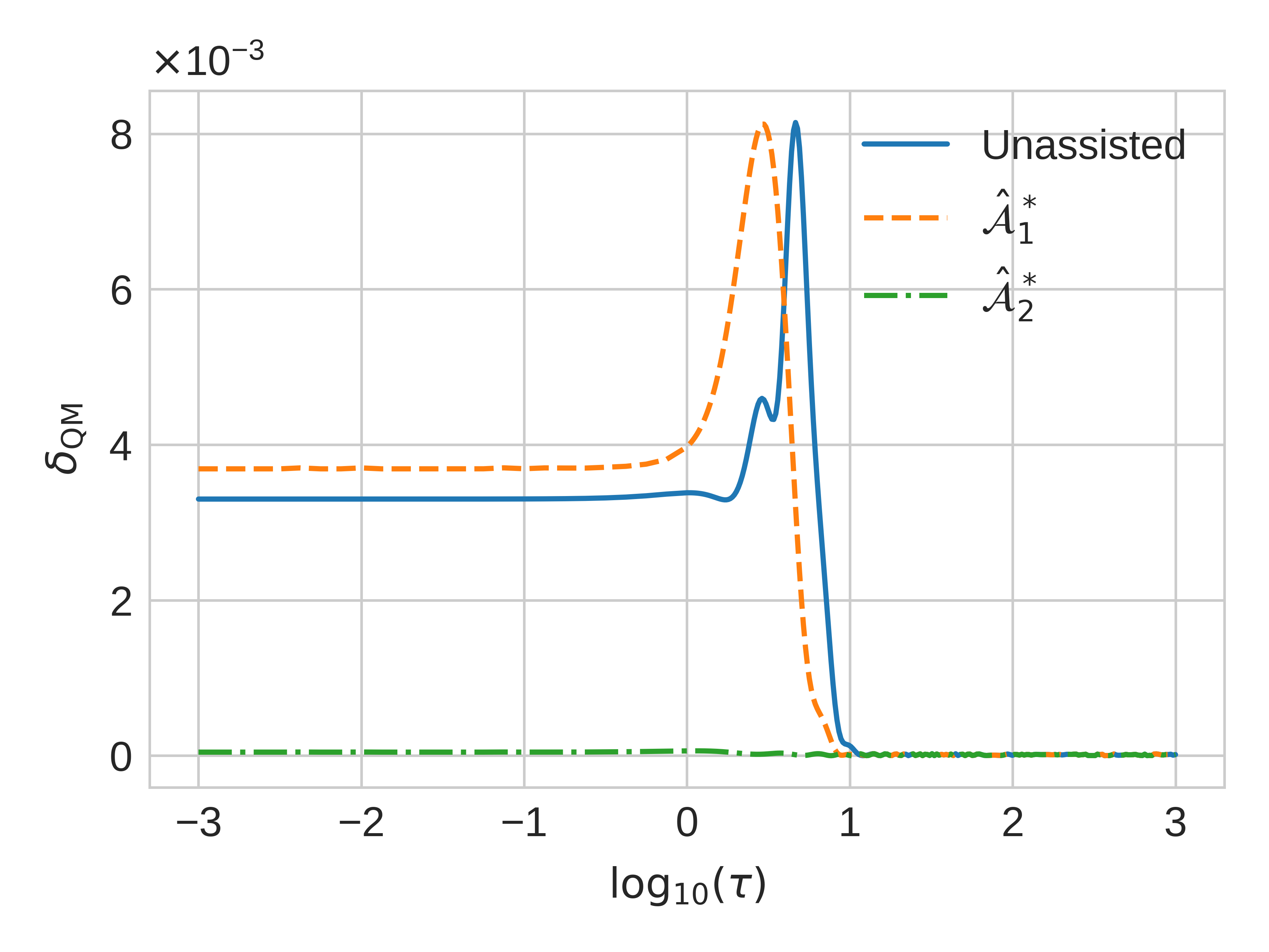}
		\caption{}
		\label{fig:qnlo tvar}
	\end{subfigure}

	\caption{Candidate FoMs in a quantum anharmonic oscillator. Note that all three FoMs agree only when the suppression is dramatic (second-order driving). The FoMs do not agree on the performance of first-order driving.}
	\label{fig:qnlo foms}
\end{figure}

\begin{figure}
	\centering
	\includegraphics[width=8.6cm]{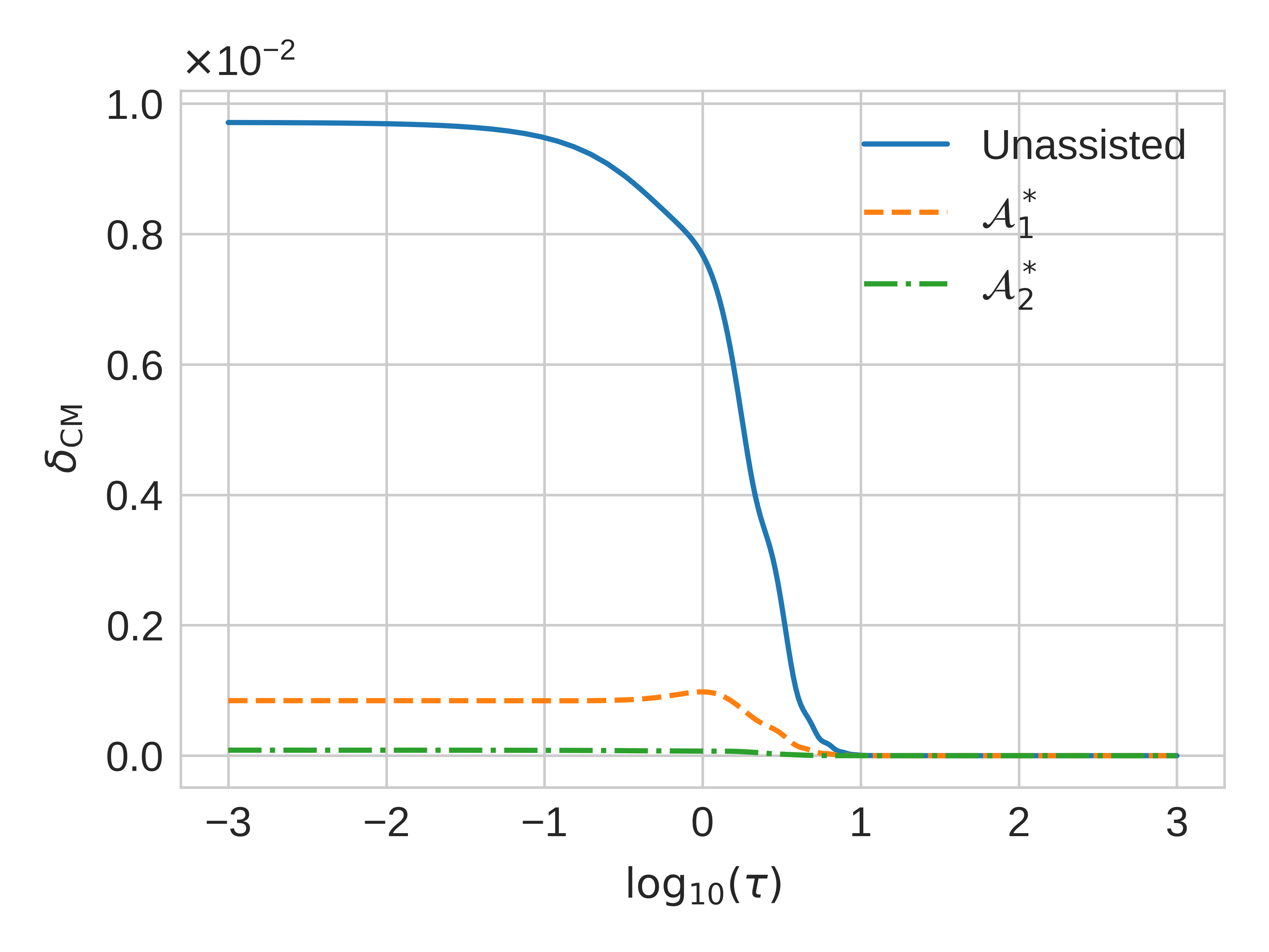}
	\caption{Classical definition of temporal variance vs. turn-on time for the single harmonic oscillator in Sec. \ref{sec:nlo}. Note that these curves closely match those in Fig.~\ref{fig:cnlo statevar}.}
	\label{fig:cnlo tvar}
\end{figure}

In addition to these two FoMs, we also considered a third candidate: temporal variance. If our transport isn't adiabatic and $\ket{\psi_f}$ is not a steady state, a general observable $\mathcal{O}$ will have time-dependent matrix elements. As long as $\mathcal{O}$ is not conserved (like total energy), we can take its expectation value $\langle \mathcal{O} \rangle (t)$ and find its infinite-time variance: $\overline{\langle \mathcal{O} \rangle^2} - \overline{\langle \mathcal{O} \rangle}^2$, where the overline denotes time averaging over $t \in [\tau, \infty)$. For our single anharmonic oscillator, we chose the observable to be the harmonic part of the Hamiltonian, such that our temporal variance became
\begin{align}
	\delta_{\mathrm{QM}} \equiv \overline{\langle H_{\mathrm{lin}} \rangle^2} - \overline{\langle H_{\mathrm{lin}} \rangle}^2, \quad t \in [\tau, \infty) ,
\end{align}
where $H_{\mathrm{lin}} = (p^2 + x^2)/2$. With these definitions, we can now simulate them in the quantum mechanical version of our single anharmonic oscillator. Standard substitutions between quantum and classical apply: functions become operators, averages become expectation values, and Poisson brackets become commutators up to a factor of $i \hbar$. Because in quantum systems position and momentum operators do not commute, we use symmetrized (Weyl) ordering of the quantum operators in the variational ansatz (see e.g. Ref.~\cite{Claeys_Eigenstates_2021} for details):
\begin{align}
	\mathcal{A}_{\beta}^{(1)} = \gamma_1 x^3 p \quad \rightarrow \quad \hat{\mathcal{A}}_{\beta}^{(1)} = \frac{\gamma_1}{2} \left( \hat{x}^3 \hat{p} + \hat{p} \hat{x}^3 \right) ,
\end{align}
where we have added hats to emphasize the difference between functions and operators. Similarly for the second-order ansatz we use:
\begin{align}
	\hat{\mathcal{A}}_{\beta}^{(2)} = \frac{\gamma_1}{2} \left( \hat{x}^3 \hat{p} + \hat{p} \hat{x}^3 \right) + \frac{\gamma_2}{4} \left( \hat{x} \hat{p}^3 + \hat{p} \hat{x} \hat{p}^2 + \hat{p}^2 \hat{x} \hat{p} + \hat{p}^3 \hat{x} \right) .
\end{align}

We can now optimize these ansatzes and test our candidate FoMs. The results are shown in Fig. \ref{fig:qnlo foms}. First, consider unassisted driving: all 3 FoMs interpolate between a quench plateau and zero in the adiabatic limit. However, both fidelity and temporal variance develop peaks at intermediate turn-on times, which likely emerge from some resonant process in the system. Next, look to first-order ACD driving. With the fidelity FoM, this ansatz only suppresses transitions at intermediate times relative to no driving; it has essentially no effect in the $\tau\to 0$ limit. Temporal energy fluctuations even slightly increase at short first-order ACD protocols compared to unassisted driving. Overall this analysis suggests that while first-order ACD suppresses energy fluctuations, it does not really bring the system closer to the adiabatic limit at short protocol times. The situation changes dramatically if we analyze the second-order ACD driving where now all three FoMs show dramatic improvement over the unassisted protocol. We conclude that both in quantum and in classical systems significant suppression of diabatic transitions comes from the presence of the cubic in momentum counter-term.

We can now move to the classical anharmonic oscillator. In this limit, the fidelity FoM obviously won't survive, but we also have to modify our temporal variance FoM. In the quantum case, the discrete spectrum allowed us to take an infinite-time variance and have a non-zero, sensible answer. However, when we move to the classical limit, our spectrum becomes continuous. This fact means that for a distribution with non-zero spread in energy, the infinite-time average of temporal fluctuations will vanish. Although at short times, observables will oscillate, the continuum of frequencies will cause the system to dephase and relax to a steady state. To construct a sensible FoM, we must focus our attention on this period of transient oscillations before relaxation, and as such, we need some characteristic time to average over. 


Because this behavior arises from energy spread, we take our characteristic time to be
\begin{align}
	T_{\mathrm{char}} \equiv \frac{E_0}{\omega \sigma_E} ,
\end{align}
where $\sigma_E$ is the standard deviation in energy over the distribution, measured immediately after a quench turn-on. Because $E_0 = 1$ and $\omega=1$ for our system, the characteristic time is just the inverse of this deviation. We can now redefine our temporal variance FoM as
\begin{align}
	\delta_{\mathrm{CM}} \equiv \overline{\langle H_{\mathrm{lin}} \rangle^2} - \overline{\langle H_{\mathrm{lin}} \rangle}^2, \quad t \in [\tau, \tau + T_{\mathrm{char}}] ,
\end{align}
where $H_{\mathrm{lin}}$ is the classical version of the harmonic Hamiltonian, and the angle brackets now indicate a classical average over our probability distribution.

In Fig. \ref{fig:cnlo tvar} we plot our classical temporal variance as a function of turn-on time for the same simulation as in Fig. \ref{fig:cnlo statevar}. Comparing the two figures, we see that both classical FoMs essentially agree on performance, although their turn-on curves have slightly different shapes.

We find that for the analyzed $\beta$-FPUT problem, energy variance is a preferable FoM because it is less ambiguous than $\delta_{\mathrm{CM}}$ and does not require choosing some rather arbitrary energy window. However, we point out that in other situations temporal fluctuations might be preferable. For example, it is easy to check that for our initial conditions in the $\alpha$-FPUT model the energy variance remains zero even in the quench limit because the cubic nonlinear term identically vanishes. Even in the $\beta$ model, temporal fluctuations might be also easier to detect experimentally, so the corresponding FoM is easier to measure.

\section{Long Wavelength Sensitivity}
\label{app:sensitivity}

In this appendix, we present evidence for why the long wavelength instability in the text seems to arise because our system (and FoM) is very sensitive to changes in $\gamma_i$ rather than the $\gamma_i$s being sensitive to changes in energy or phase space manifold. Although this analysis is by no means exhaustive, it does function as a preliminary treatment to be improved upon in future work.

\subsection*{Optimization sensitivity}

Our first goal is to investigate the sensitivity of $\gamma_i(\beta)$ to changes in our chosen region of phase space. We will quantify this ``region'' by choosing different values of $d_E$ corresponding to different widths of our microcanonical distribution. In order to quantify the sensitivity of the coefficients, we will compute the functional overlap of $\gamma_i(\beta)$ for some $d_E$ in question with $\gamma_i(\beta)$ for $d_E = 0$. More specifically, we use
\begin{align}
	F^2_i(d_E) = \frac{\left(\int_{0}^{1} \gamma_i(\beta,d_E) \gamma_i(\beta,d_E=0) d\beta\right)^2}{\left( \int_{0}^{1} \gamma_i^2(\beta,d_E) d\beta \right) \left( \int_{0}^{1} \gamma_i^2(\beta,d_E=0) d\beta \right)} .
\end{align}
If the optimization process is not sensitive to small changes in $d_E$, we would expect this overlap to smoothly interpolate between $1$ and some value $1 - \epsilon$ as our width increases from 0, indicating that the coefficients are not very sensitive to $d_E$. However, if they are sensitive, we would expect a sharp drop in the overlap to some value near zero.

\begin{figure}
	\centering
	\includegraphics[width=8.6cm]{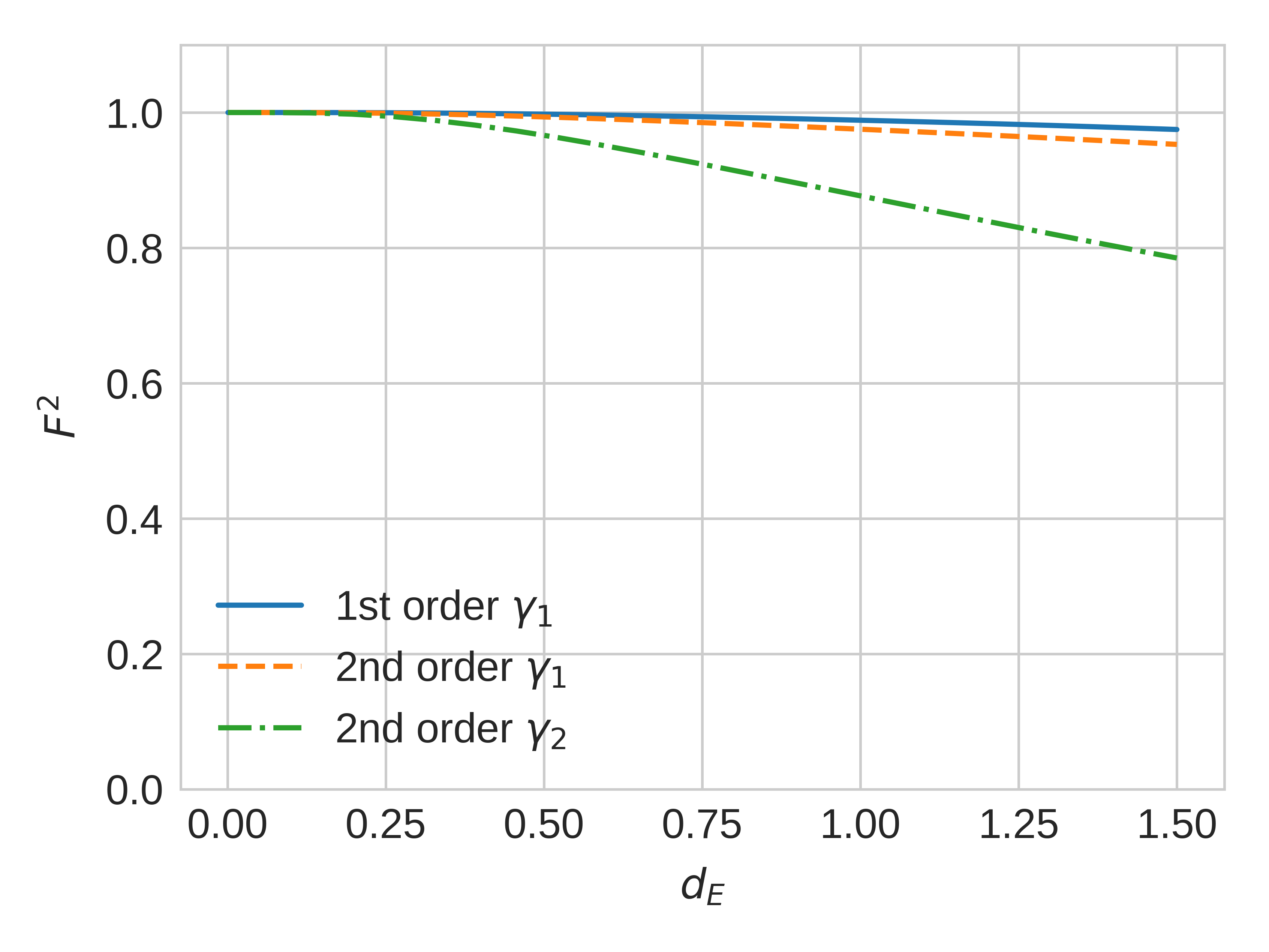}
	\caption{Functional overlap of driving coefficents $\gamma_i(\beta)$ at width $d_E$ with same coefficients at width $d_E = 0$. The values interpolate smoothly and remain close to $1$ by the time $d_E = 0.4$, indicating that the instability does not appear during optimization.}
	\label{fig:bfput5 gamma overlap}
\end{figure}

In Fig. \ref{fig:bfput5 gamma overlap}, we plot this metric for $d_E \in [0,1.5]$ in our long-wavelength $N=5$ setup. As is evident from the plot, $F^2$ interpolates smoothly as we vary the width, and by the time $d_E = 0.4$, the lowest overlap is still around $0.97$. As such, we can conclude that the sensitivity of the ACD protocol is not coming from unstable variational minimization procedure.

\subsection*{System sensitivity}

\begin{figure}
	\centering
	\includegraphics[width=8.6cm]{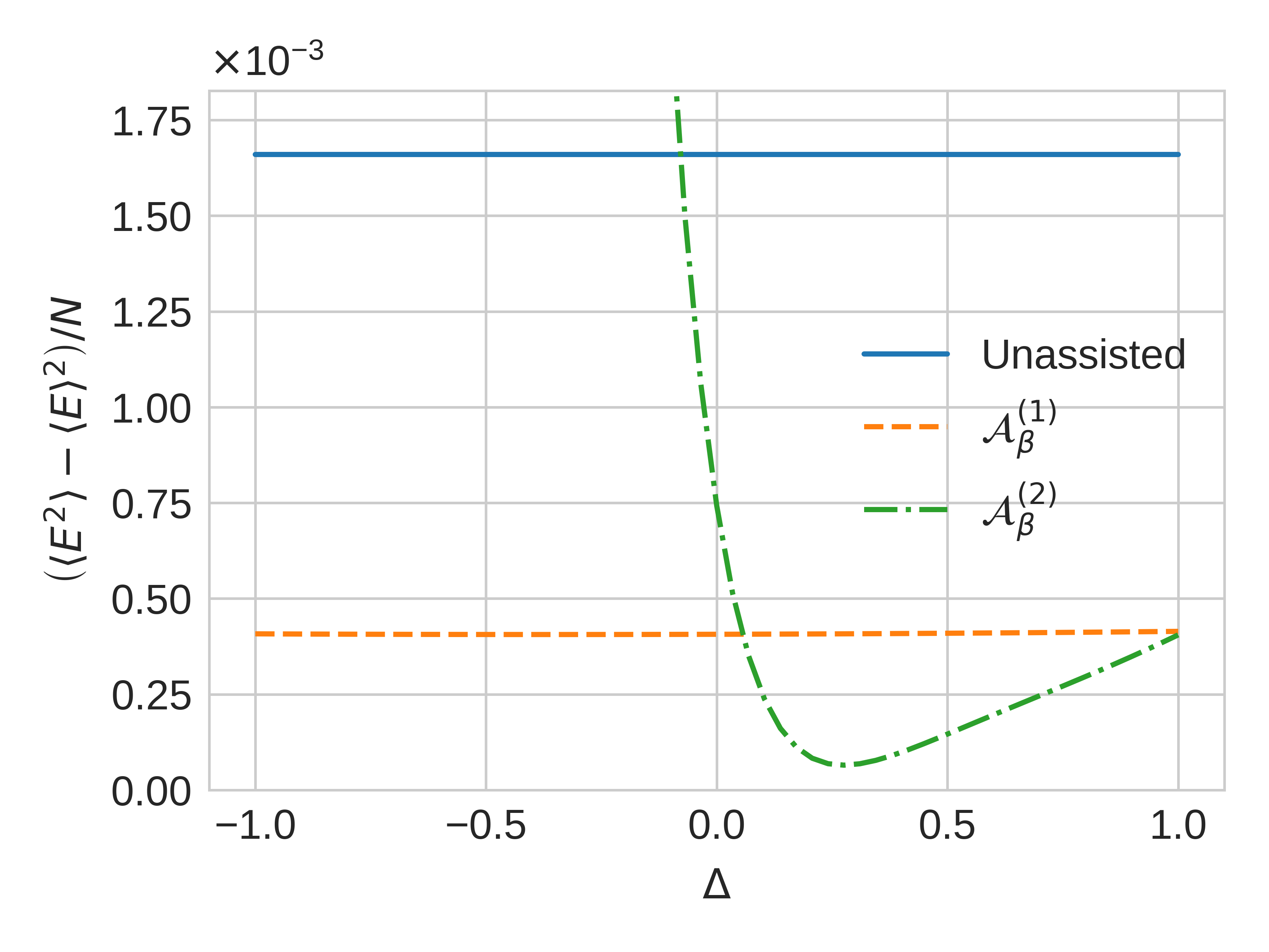}
	\caption{Final energy variance (over system size) after quench turn-on ($\tau \approx 3\cdot 10^{-4}$) as a function of perturbation $\Delta$. We perform the simulations in the $N=5$ $\beta$-FPUT chain with $E_0 = 1$ and $d_E = 0$.}
	\label{fig:bfput5 Delta SS}
\end{figure}

\begin{figure}
	\centering
	\includegraphics[width=8.6cm]{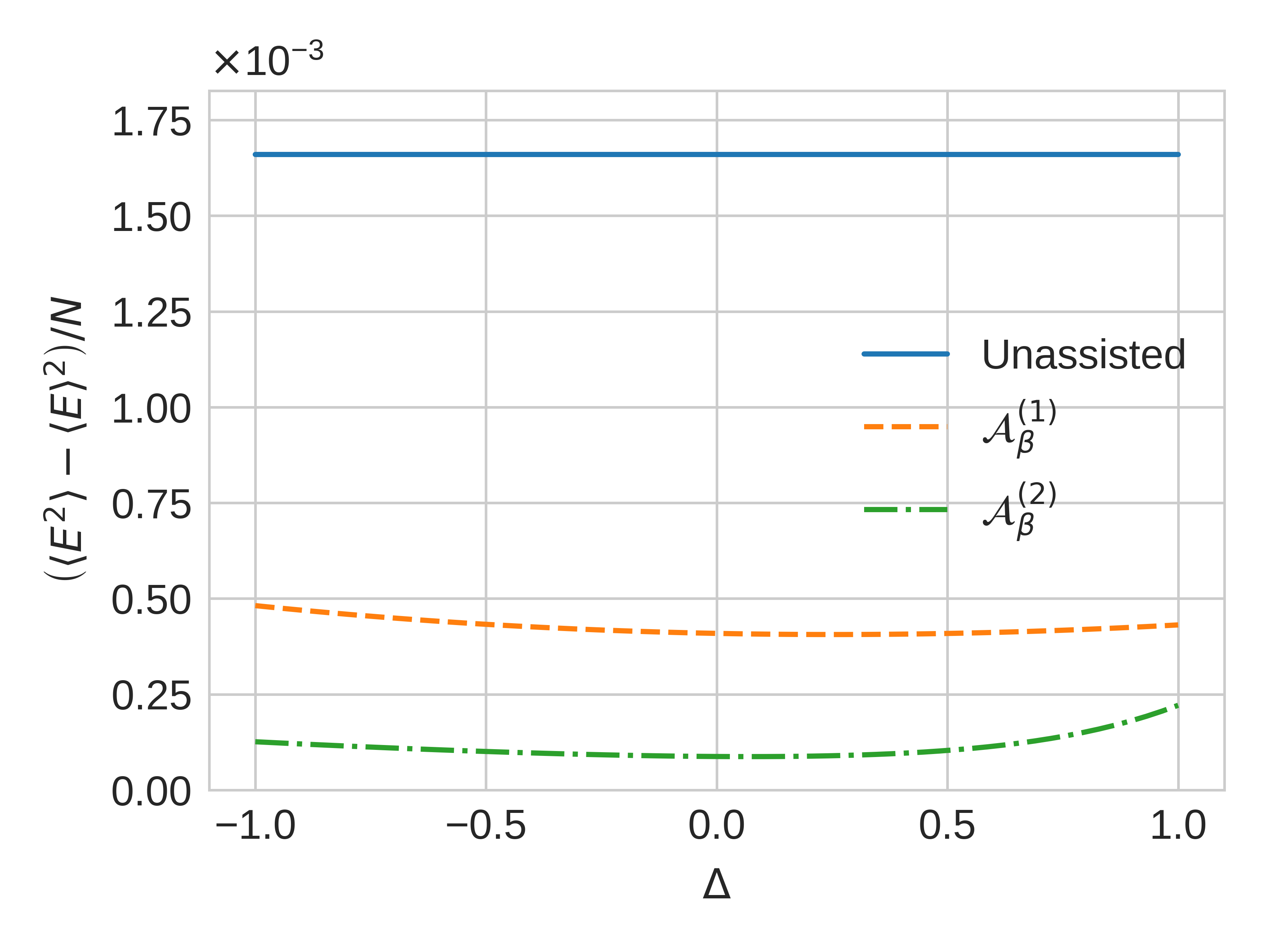}
	\caption{Final energy variance (over system size) after quench turn-on ($\tau \approx 3\cdot 10^{-4}$) as a function of perturbation $\Delta$. We perform the simulations in the $N=5$ $\beta$-FPUT chain with $E_0 = 1$ and $d_E = 0.4$.}
	\label{fig:bfput5 Delta G}
\end{figure}

Our next goal is to investigate the sensitivity of our energy variance FoM to small changes in our driving parameters $\gamma_i(\beta)$. To simplify the process of adding perturbations to these functional forms, we will use the following method: first, choose a perturbation strength $\Delta$. Then, to each of the optimized coefficients, perturb the rational fit from Eq.~\eqref{gamma ansatz} such that it becomes
\begin{align}
	\gamma_i'(\beta) = \frac{b_0 + b_1 (1 + \Delta) \beta + b_2 \beta^2 + b_3 \beta^3 + \cdots}{1 + c_1 \beta + c_2 \beta^2 + c_3 \beta^3 + \cdots} .
	\label{Delta pert}
\end{align}
Now, simulate a quench turn-on (sufficiently small $\tau$) and measure the energy variance FoM. Repeat for many different values of $\Delta$. The dependence of the FoM on this perturbation will reveal how sensitive the system is to small changes in the driving.

In Fig. \ref{fig:bfput5 Delta SS}, we plot the FoM as a function of $\Delta$ for the case where our width is $d_E = 0$. We use a quench turn-on time of $\tau \approx 3\cdot 10^{-4}$ and the same $E_0 = 1$, so just as expected the $\Delta = 0$ line corresponds to the far left of Fig. \ref{fig:bfput5 statevar}. We see that near $\Delta = 0$, the curve is essentially flat, indicating that the system is not very sensitive to perturbations in this driving. Meanwhile, the curve for second-order has a large negative slope when $\Delta = 0$, indicating a strong dependence on the driving coefficients. It is here that we observe the sensitivity necessary to explain the long-wavelength instability, not in the optimization process.

We can also perform the same analysis when optimizing for width $d_E = 0.4$, the results of which are plotted in Fig. \ref{fig:bfput5 Delta G}. Now, both curves are essentially flat in the neighborhood of $\Delta = 0$. These attributes indicate that our method of fighting the instability actually worked: exploring a larger neighborhood of phase space suppressed the system's sensitivity to driving coefficients.



%

\end{document}